\documentclass[a4paper,11pt]{article}

\usepackage{color}
\usepackage{latexsym}
\usepackage{amssymb}
\usepackage{amsthm}
\usepackage{amsmath}
\usepackage{amsfonts}
\usepackage{wrapfig}
\usepackage[dvipdfmx]{graphicx}

\newcommand{\eqb}{\begin{equation}}
\newcommand{\eqe}{\end{equation}}
\newcommand{\eqbnon}{\begin{equation*}}
\newcommand{\eqenon}{\end{equation*}}

\newcommand{\eqab}{\begin{eqnarray}}
\newcommand{\eqae}{\end{eqnarray}}
\newcommand{\eqabnon}{\begin{eqnarray*}}
\newcommand{\eqaenon}{\end{eqnarray*}}

\newcommand{\seqb}{\begin{subequations}}
\newcommand{\seqe}{\end{subequations}}

\newcommand{\eref}[1]{\eqref{#1}}

\newcommand{\defeq}{:=}

\newcommand{\od}[2]{\dfrac{{\rm d} #1}{{\rm d} #2}}
\newcommand{\pd}[2]{\dfrac{\partial #1}{\partial #2}}

\newcommand{\odline}[2]{{\rm d} #1/{\rm d} #2}
\newcommand{\pdline}[2]{\partial #1/\partial #2}
\newcommand{\ls}[1]{_{\rm #1}}

\newcommand{\ld}[1]{\overset{\bullet}{#1}}

\newcommand{\Teq}{T\ls{eq}}
\newcommand{\eeq}{\varepsilon\ls{eq}}
\newcommand{\peq}{p\ls{eq}}
\newcommand{\seq}{s\ls{eq}}
\newcommand{\rhoeq}{\rho\ls{eq}}
\newcommand{\Veq}{V\ls{eq}}

\newcommand{\Tne}{T\ls{ne}}
\newcommand{\ene}{\varepsilon\ls{ne}}
\newcommand{\pne}{p\ls{ne}}
\newcommand{\sne}{s\ls{ne}}
\newcommand{\rhone}{\rho\ls{ne}}
\newcommand{\Vne}{V\ls{ne}}

\newcommand{\sv}{\overset{\circ}{\Pi}}

\newcommand{\tauh}{\tau\ls{h}}
\newcommand{\taub}{\tau\ls{b}}
\newcommand{\taus}{\tau\ls{s}}

\newcommand{\inthb}{\beta\ls{hb}}
\newcommand{\inths}{\beta\ls{hs}}

\newcommand{\ghb}{\gamma\ls{hb}}
\newcommand{\ghs}{\gamma\ls{hs}}

\newcommand{\tssp}[1]{[\![\, #1 \,]\!]^{\circ}}

\newtheorem{ass}{Assumption}
\newtheorem{con}{Supplemental Condition}

\begin{document}
%%%%%%%%%%%%%%%%%%%%%%%%%%%%%%%%%%%%%%%%%%%%%%%%%%%%%%%%%%%%%%%%%%%%%%%%%%%%%%%%%%%%%%%%%%%%%%%%%%%%

\title{
An Axiomatic Review of Israel-Stewart Hydrodynamics and Extended Irreversible Thermodynamics
}

\author{
Hiromi Saida (Daido University, Japan)
}
\date{
%{\small email: saida@daido-it.ac.jp}
}

\maketitle

\abstract{\small
The causality of dissipative phenomena can not be treated in traditional theories of dissipations, Fourier laws and Navier-Stokes equations. 
This is the reason why the dissipative phenomena have not been studies well in relativistic situations. 
Furthermore, the interactions among dissipations, e.g. the heating of fluid due to viscous flow and the occurrence of viscous flow due to heat flux, are not explicitly described in those traditional laws. 
One of the phenomenologies which describe the causality and interaction of dissipations is the Extended Irreversible Thermodynamics (EIT). 
(Israel-Stewart theory of dissipative hydrodynamics is one approximate form of EIT.) 
This manuscript reviews an axiomatic construction of EIT and Israel-Stewart hydrodynamic theory. 
Also, we point out that the EIT is also applicable to radiative transfer in optically \emph{thick} matters. 
However, radiative transfer in optically \emph{thin} matters can not be described by EIT, because the non-self-interacting nature of photons is incompatible with a basic requirement of EIT, ``the bilinear form of entropy production rate''. 
The break down of EIT in optically thin situation is not explicitly recognized in standard references of EIT and Israel-Stewart theory. 
Some detail of how EIT fails to describe a radiative transfer in optically thin situations is also explained. 
(This manuscript is a revision of the contribution to a book~\cite{ref:book} published in 2011. 
So, recent developments made after 2011 may not be cited.)
}

%%%%%%%%%%%%%%%%%%%%%%%%%%%%%%%%%%%%%%%%%%%%%%%%%%%%%%%%%%%%%%%%%%%%%%%%%%%%%%%%%%%%%%%%%%%%%%%%%%%%
\section{Introduction}
\label{sec:intro}

As a physically rigorous condition, the causality has to be preserved in any phenomena. 
However, it is hard to make the theories of dissipations preserve the causality. 
If one relies on the Navier-Stokes and Fourier laws which we call \emph{traditional laws of dissipations}, then an infinite speed of propagation of dissipations is concluded~\cite{ref:eit_2}. (See appendix~\ref{app:mc} for a short summary.) 
This means the breakdown of causality, which is the reason why the dissipative phenomena have not been studies well in relativistic situations. 
Also, the infinitely fast propagation denotes that, even in non-relativistic cases, the traditional laws of dissipations can not describe dynamical behaviors of fluid whose dynamical time scale is comparable with the time scale within which non-stationary dissipations relax to stationary ones.

Moreover note that, since Navier-Stokes and Fourier laws are independent phenomenological laws, interaction among dissipations, e.g. the heating of fluid due to viscous flow and the occurrence of viscous flow due to heat flux, are not explicitly described in those traditional laws. 
(See appendix~\ref{app:mc} for a short summary.) 
Thus, in order to find a physically reasonable theory of dissipative fluids, it is expected that not only the finite speed of propagation of dissipations but also the interaction among dissipations are included in the desired theory of dissipative fluids.

Problems of the infinite speed of propagation and the absence of interaction among dissipations can be resolved if we rely not on the traditional laws of dissipations but on the \emph{Extended Irreversible Thermodynamics} (EIT)~\cite{ref:eit_1,ref:eit_2}. 
The EIT, both in non-relativistic and relativistic situations, is a causally consistent phenomenology of dissipative fluids including interaction among dissipations~\cite{ref:hl.causality}.
Note that the non-relativistic EIT may have some experimental grounds for laboratory systems~\cite{ref:eit_2}. 
Thus, although an observational or experimental verification of relativistic EIT has not been obtained so far, the EIT may be one of physically preferable hydrodynamic theories for dissipative fluids even in relativistic situations. 
Here, one may refer to the relativistic hydrodynamics proposed originally by Israel~\cite{ref:is_1} and developed by Israel and Stewart~\cite{ref:is_2}, which describes the causal propagation of dissipations. 
However, since the Israel-Stewart hydrodynamics can be derived as an approximate form of EIT (see Sec.\ref{sec:eqs}), we dare to use the term EIT in this review manuscript.

In astrophysics, recent advances in technology of astronomical observation is going to realize a fine observation whose resolution is close to the view size of black hole candidates~\cite{ref:obs_3,ref:obs_1,ref:obs_2}. 
(Note that, at present in electro-magnetic wave observations, no certain/direct evidence of the existence of black hole has been extracted from observational data.) 
The light that is to be detected by our telescope is emitted by the matter accreting on to a black hole, and the energy of the light is supplied via the dissipations in accreting matter. 
Therefore, observational data of black hole candidates should include signals of dissipative phenomena in strong gravitational field around black hole. 
We expect that such general relativistic dissipative phenomena is described by the EIT.

EIT has been used to consider some phenomena in the presence of strong gravity. 
For example, Peitz and Appl~\cite{ref:pa} have used EIT to write down a set of evolution equations of dissipative fluid and spacetime metric (gravitational field) for stationary axisymmetric situation. 
However the Peitz-Appl formulation looks very complicated, and has predicted no concrete result on astrophysics so far. 
Another example is the application of EIT to a dissipative gravitational collapse under the spherical symmetry. 
Herrera and co-workers~\cite{ref:collapse_1,ref:collapse_2,ref:collapse_3} constructed some models of dissipative gravitational collapse with some simplification assumptions. 
They rearranged the basic equations of EIT into a suitable form, and deduced some interesting physical implications about dissipative gravitational collapse. 
But, at present, there still remain some complexity in Herrera's system of equations for gravitational collapse, and it seems not to be applicable to the understanding of observational data of black hole candidates~\cite{ref:obs_3,ref:obs_1,ref:obs_2}. 
These facts imply that, in order to extract the signals of strong gravity from the observational data of black hole candidates, we need a more sophisticated strategy for the application of EIT to general relativistic dissipative phenomena. 
In order to construct the sophisticated strategy, we may need to understand the EIT deeply.

Then, this review manuscript aims to show an axiomatic understanding of EIT. 
We expect that the axiomatic formulation can clarify the basic ideas and applicable range of EIT. 
We try to understand the EIT from the point of view of non-equilibrium physics, because the EIT is regarded as dissipative hydrodynamics based on the idea that the thermodynamic state of each fluid element is a non-equilibrium state. 
(But thorough knowledge of non-equilibrium thermodynamics and general relativity is not needed in reading this manuscript.) 
As explained in detail in following sections, the non-equilibrium nature of fluid element arises from the dissipations which are essentially \emph{irreversible processes}. 
Then, non-equilibrium thermodynamics applicable to each fluid element is constructed in the framework of EIT, which includes the interaction among dissipations and describes the causal entropy production process due to the dissipations. 
Furthermore we point out that the EIT is applicable also to radiative transfer in optically \emph{thick} matters~\cite{ref:rad.dense_2,ref:rad.dense_1}. 
However, radiative transfer in optically \emph{thin} matters can not be described by EIT, because the non-self-interacting nature of photons is incompatible with a basic requirement of EIT. 
This is not explicitly recognized in standard references of EIT~\cite{ref:hl.causality,ref:hl.applicable,ref:eit_1,ref:eit_2,ref:is_1,ref:is_2}.

Here let us make two comments: 
Firstly, note that the EIT can be formulated with including not only heat and viscosities but also electric current, chemical reaction and diffusion in multi-component fluids~\cite{ref:eit_1,ref:eit_2}. 
Including all of them raises an inessential mathematical confusion in our discussions. 
Therefore, for simplicity of discussions in this manuscript, we consider the \emph{simple dissipative fluid}, which is electrically neutral and chemically inert single-component dissipative fluid. 
This means to consider the heat flux, bulk viscosity and shear viscosity as the dissipations in fluid.

As the second comment, we emphasize that the EIT is a \emph{phenomenology} in which the transport coefficients are parameters undetermined in the framework of EIT~\cite{ref:eit_1,ref:eit_2,ref:is_1}.
On the other hand, based on the Grad's 14-moment approximation method of molecular motion, Israel and Stewart~\cite{ref:is_2} have obtained the transport coefficients of EIT as functions of thermodynamic variables. 
The Israel-Stewart's transport coefficients are applicable to the molecular kinematic viscosity. 
However, it is not clear at present whether those coefficients are applicable to other mechanisms of dissipations such as fluid turbulent viscosity and the so-called magneto-rotational-instability (MRI) which are usually considered as the origin of viscosities in accretion flows onto celestial objects~\cite{ref:accretion_1,ref:accretion_2,ref:ss}. 
Concerning the MRI, an analysis by Pessah, Chan, and Psaltis~\cite{ref:mri_1,ref:mri_2} seems to imply that the dissipative effects due to MRI-driven turbulence can be expressed as some transport coefficients, whose form may be different from Israel-Stewart's transport coefficients. 
Hence, in this manuscript, we do not refer to the Israel-Stewart's coefficients. 
We re-formulate the EIT simply as the phenomenology, and the transport coefficients are the parameters that have to be determined empirically through observations or by underlying fundamental theories of turbulence and/or molecular dynamics. 
The determination of transport coefficients and the investigation of micro-processes of transport phenomena are out of the aim of this manuscript. 
The point of EIT in this manuscript is the causality of dissipations and the interaction among dissipations.

In Sec.\ref{sec:ass}, the basic ideas of EIT is summarized into four assumptions and one supplemental condition, and a limit of EIT is also reviewed. 
Sec.\ref{sec:eqs} explains the meanings of basic quantities and equations of EIT, and also the derivation of basic equations are summarized so as to be extendable to fluids which are more complicated than the simple dissipative fluid. 
Sec.\ref{sec:rad} is for a remark on non-equilibrium radiative transfer, of which the standard references of EIT were not aware. 
Sec.\ref{sec:conc} gives a concluding remark on a tacit understanding which is common to EIT and traditional laws of dissipations.

Throughout this manuscript, the semicolon ``~$;$~'' denotes the covariant derivative with respect to spacetime metric, while the comma ``~$,$~'' denotes the partial derivative. 
The definition of covariant derivative is summarized in appendix~\ref{app:cd}. 
(Thorough knowledge of general relativity is not needed in reading this manuscript, but experiences of calculation in special relativity is preferable.) 
The unit used throughout is $c = 1$, $G = 1$ and $k_B = 1$. 
Since the quantum mechanics is not used, we do not care about the Planck constant.

%%%%%%%%%%%%%%%%%%%%%%%%%%%%%%%%%%%%%%%%%%%%%%%%%%%%%%%%%%%%%%%%%%%%%%%%%%%%%%%%%%%%%%%%%%%%%%%%%%%%
\section{Basic assumptions and a supplemental condition of EIT}
\label{sec:ass}

We begin with summarizing the basis of perfect fluid and traditional laws of dissipations. 
The theory of perfect fluid is a phenomenology assuming the \emph{local (thermal) equilibrium}; each fluid element is in a thermal equilibrium state. 
Note that, exactly speaking, dissipations can not exist in thermal equilibrium states. 
Thus the local equilibrium assumption is incompatible with the dissipative phenomena which are essentially the irreversible and entropy producing processes. 
Due to the local equilibrium assumption, the basic equations of perfect fluid do not include any dissipation, and each fluid element in perfect fluid evolves adiabatically. 
No entropy production arises in the fluid element of perfect fluid~\cite{ref:ll}. 
Furthermore, recall that the traditional laws of dissipations (Navier-Stokes and Fourier laws) are also the phenomenologies assuming the local equilibrium. 
Therefore, the traditional laws of dissipations lead inevitably some unphysical conclusions, one of which is the infinitely fast propagation of dissipations~\cite{ref:eit_1,ref:eit_2}.

From the above, it is recognized that we should replace the local equilibrium assumption with the idea of \emph{local \underline{non}-equilibrium} in order to obtain a physically consistent theory of dissipative phenomena. 
This means to consider that the fluid element is in a non-equilibrium thermodynamic state.
A phenomenology of dissipative irreversible hydrodynamics, under the local \emph{non}-equilibrium assumption, is called the \emph{Extended Irreversible Thermodynamics} (EIT).~\footnote{
Although the EIT is a dissipative ``hydrodynamics'', it is named ``thermodynamics''. 
This name puts emphasis on the replacement of local equilibrium idea with local non-equilibrium one, which is a revolution in thermodynamic treatment of fluid elements.}
In this manuscript, we summarize the basic ideas of EIT into four assumptions and one supplemental condition. 
And, as mentioned in Sec.\ref{sec:intro}, we consider the simple dissipative fluid in which the dissipations are heat flow, bulk viscosity and shear viscosity.

The first assumption of EIT is as follows:
\begin{ass}[Local Non-equilibrium]
The dissipative fluid under consideration is in ``local'' non-equilibrium thermodynamic states. 
This means that each fluid element is in a non-equilibrium state, but the non-equilibrium state of one fluid element is not necessarily the same with the non-equilibrium state of the other fluid element.~$\clubsuit$
\end{ass}

Due to this assumption, it is necessary for the EIT to formulate a non-equilibrium thermodynamics to describe thermodynamic state of each fluid element. 
In order to formulate it, we must specify the state variables which are suitable for characterizing non-equilibrium states. 
The second assumption of EIT is on the specification of suitable state variables for non-equilibrium states of fluid elements:
\begin{ass}[Non-equilibrium thermodynamic state variables]
The state variables which characterize the non-equilibrium states are distinguished into two categories;
\begin{description}
\item[1st category (Traditional Variables)] The state variables in this category do not necessarily vanish at the local equilibrium limit of fluid. 
These are the variables specified already in equilibrium thermodynamics, e.g. the temperature, internal energy, pressure, entropy and so on. 
\item[2nd category (Dissipative Fluxes)] The state variables in this category should vanish at the local equilibrium limit of fluid. 
These are, in the framework of EIT of simple dissipative fluid, the ``heat flux'', ``bulk viscosity'', ``shear viscosity'' and their thermodynamic conjugate state variables. 
(e.g. thermodynamic conjugate to entropy $S$ is temperature $T \equiv \partial E/\partial S$, where $E$ is internal energy. 
Similarly, thermodynamic conjugate to bulk viscosity $\Pi$ can be given by $\partial E/\partial \Pi$, where $E$ is now ``non-equilibrium'' internal energy.)~$\clubsuit$
\end{description}
\end{ass}

Note that the terminology ``traditional variables'' is a coined word introduced by the present author, and not a common word in the study on EIT. 
But let us dare to use the term ``traditional variables'' to explain clearly the idea of EIT.

Next, recall that, in the ordinary equilibrium thermodynamics, the number of independent state variables is two for closed systems which conserve the number of constituent particles, and three for open systems in which the number of constituent particles changes. 
For non-equilibrium states of dissipative fluid elements, it seems to be natural that the number of independent traditional variables is the same with that of state variables in ordinary equilibrium thermodynamics. 
On the other hand, in the traditional laws of dissipations which are summarized in Eq.\eref{eq:mc.classic} in appendix~\ref{app:mc}, the dissipative fluxes such as heat flux and viscosities were not independent variables, but some functions of fluid velocity and local equilibrium state variables such as temperature and pressure. 
However in the EIT, the dissipative fluxes are assumed to be independent of fluid velocity and traditional variables. 
Then, the third assumption of EIT is on the number of independent state variables:
\begin{ass}[Number of Independent State Variables]
The number of independent traditional variables is the same with the ordinary thermodynamics (two for closed system and three for open system). 
Furthermore, EIT assumes that the number of independent dissipative fluxes for simple dissipative fluid are three. 
For example, we can regard the heat flux, bulk viscosity and shear viscosity are independent dissipative fluxes.~{\rm $\clubsuit$~\footnote{As an advanced remark, recall that, in ordinary equilibrium thermodynamics, if there is an external field such as a magnetic field applied on a magnetized gas, then the number of independent state variables increases for both closed and open systems. 
Usually, the external field itself can be regarded as an additional independent state variable. 
The same is true of the number of independent traditional variables. 
In this manuscript, we consider no external field other than the external gravity (the spacetime metric tensor), and the metric can be regarded as an additional independent traditional variable. 
However, for simplicity and in order to focus our attention to \emph{intrinsic} state variables of non-equilibrium states, we do not explicitly show the metric as an independent state variable in all discussions in this manuscript.}}
\end{ass}

Mathematically, the independent state variables can be regarded as a ``coordinate system'' in the space which consists of thermodynamic states. 
A set of values of the coordinates corresponds to a particular non-equilibrium state. 
This means that the set of values of independent state variables is uniquely determined for each non-equilibrium state, and different non-equilibrium states have different sets of values of independent state variables.

Assumption~3 implies the existence of non-equilibrium equations of state, from which we can obtain ``dependent'' state variables. 
Concrete forms of non-equilibrium equations of state should be determined by experiments or micro-scopic theories of dissipative fluids. 
Suppose that non-equilibrium equations of state are given, then we can consider a case that the non-equilibrium entropy, $S\ls{ne}$, is chosen as a dependent state variable. 
In this case, $S\ls{ne}$ depends on an independent dissipative flux, e.g. the bulk viscosity $\Pi$. 
Obviously, since $\Pi$ is one of the dissipative fluxes, the partial derivative, $\partial S\ls{ne}/\partial \Pi$, has no counter-part in ordinary equilibrium thermodynamics. 
This implies that $\partial S\ls{ne}/\partial \Pi$ is a member not of traditional variables but of dissipative fluxes. 
Hence, by the assumption~2, we find that the partial derivative of dependent state variable by an independent dissipative flux should vanish at local equilibrium limit of fluid,
\eqb
\label{eq:eit.ass3}
 \pd{[\mbox{\rm non-equilibrium dependent state variable}]}
    {[\mbox{\rm independent dissipative flux}]} \to 0
  \quad \mbox{as {\rm [independent dissipative fluxes]}} \to 0 \,.
\eqe

The assumptions~1, 2 and~3 can be regarded as the zeroth law of non-equilibrium thermodynamics formulated in the EIT, which prescribes the existence and basic properties of local non-equilibrium states of dissipative fluids. 
In the relativistic formulation of EIT, the state variables are gathered in the energy-momentum tensor, $T^{\mu\nu}$. 
The definition of $T^{\mu\nu}$ will be shown in next section.

Note that, because EIT is a ``hydrodynamics'', we should consider not only non-equilibrium thermodynamic state variables but also a \emph{dynamical variable}, the fluid velocity. 
As will be shown in next section, the basic equations of EIT determine not only non-equilibrium state variables but also the dynamical variable (velocity) of fluid element, when initial and boundary conditions are specified. 
Via those basic equations, \emph{the fluid velocity can be regarded as a function of thermodynamic state variables of fluid elements.} 
Further note that the evolution equations of fluid velocity and traditional variables are given by the same derivation as the perfect fluid hydrodynamics, which are the conservation laws of mass current vector and energy-momentum tensor.

Those method to derive the evolution equations of fluid velocity and traditional variables cannot predict evolution equations of dissipative fluxes. 
Hence, we need guiding principles to derive the evolution equations of dissipative fluxes. 
In the EIT, the guiding principles are the second law of thermodynamics and the phenomenological requirement based on laboratory experiments summarized in appendix~\ref{app:mc}. 
The fourth assumption of EIT is on these guiding principles:
\begin{ass}[Second Law and Phenomenology]
The self-production rate of entropy by a fluid element, $\sigma\ls{s}(x)$, at spacetime point $x$ is defined by the divergence of non-equilibrium entropy current vector, $\sigma\ls{s} \defeq S\ls{ne}^{\,\,\,\,\mu}\,_{;\mu}$, where the detail of $S\ls{ne}^{\,\,\,\,\mu}$ is not necessary at present and shown in Sec.\ref{sec:eqs.derivation}. 
Concerning $\sigma\ls{s}$, EIT assumes the followings:
\begin{description}
\item[(4-a)] Entropy production rate is non-negative, $\sigma\ls{s} \ge 0$ (2nd law).
\item[(4-b)] 
Entropy production rate is expressed by the bilinear form,
\eqb
\label{eq:eit.bilinear}
 \sigma\ls{s} = \mbox{\rm [Dissipative Flux]} \times \mbox{\rm [Thermodynamic force]} \,,
\eqe
\end{description}
Here, as explained below, the ``thermodynamic force'' is given by some gradients of state variables which raises a dissipative flux. 
And, the functional form of thermodynamic force should be consistent with existing phenomenologies summarized in appendix~\ref{app:mc}. 
Concrete forms of them are derived in Sec.\ref{sec:eqs.derivation}.~$\clubsuit$
\end{ass}

The notion of thermodynamic force in requirement~(4-b) is not a particular property of EIT. 
Indeed, the thermodynamic force has already been known in traditional laws of dissipations. 
For example, the Fourier law, $\vec{q} = -\lambda \vec{\nabla}T$, implies that the temperature gradient, $-\vec{\nabla}T$, is the thermodynamic force which raises the heat flux, $\vec{q}$, where $\lambda$ is the heat conductivity.
Also, it is already known for the Fourier law that the entropy production rate is given by the bilinear form, $\vec{q}\cdot(-\vec{\nabla}T) = q^2/\lambda \ge 0$, where $\vec{q}$ and $-\vec{\nabla}T$ corresponds respectively to the dissipative flux and thermodynamic force in the above requirement~(4-b)~\cite{ref:eit_1,ref:eit_2}. 
The assumption~4 is a simple extension of the local equilibrium theory to the local non-equilibrium theory. 
The causality of dissipative phenomena is not retained by solely the assumption~4. 
Also, the inclusion of interaction among dissipative fluxes is not achieved by solely the assumption~4.

\emph{The point of preservation of causality and inclusion of interaction among dissipations is the definition of non-equilibrium entropy current, $S\ls{ne}^{\,\,\,\,\mu}$.} 
As explained in next section, once an appropriate definition of $S\ls{ne}^{\,\,\,\,\mu}$ is given, the assumption~4 together with the other three assumptions yields the evolution equations of dissipative fluxes which retain the causality of dissipative phenomena and includes the interaction among dissipative fluxes.

To find the appropriate definition of $S\ls{ne}^{\,\,\,\,\mu}$, it should be noted that, unfortunately, some critical problems have been found for the cases of strong dissipative fluxes; e.g. the uniqueness of non-equilibrium temperature, $\Tne$, and non-equilibrium pressure, $\pne$, can not be established in the present status of EIT~\cite{ref:eit_1,ref:eit_2}. 
These problems are the very difficult issues in non-equilibrium physics. 
At present, the EIT seems not to be applicable to a non-equilibrium state with strong dissipations. 
However, we can expect that, by restricting our discussion to the case of weak dissipative fluxes, the difficult problems in non-equilibrium physics is avoided and a well-defined entropy current, $S\ls{ne}^{\,\,\,\,\mu}$, is obtained. 
This expectation is realized by adopting a perturbative method summarized in the following supplemental condition:
\begin{con}[Second Order Approximation of Equations of State]
Restrict our interest to the non-equilibrium states which are not so far from equilibrium states. 
This means that the dissipative fluxes are not very strong. 
Quantitatively, we consider the cases that the strength of dissipative fluxes is limited so that the ``second order approximation of equations of state'' is appropriate: 
Examples of the second order approximation with non-equilibrium specific entropy, $\sne$, and non-equilibrium temperature, $\Tne$, are~\cite{ref:eit_1,ref:eit_2} as follows,
\seqb
\label{eq:eit.eos}
\eqab
\label{eq:eit.eos.sne}
 \sne(\ene , \Vne , q^{\mu} , \Pi , \sv\,^{\mu\nu}) &=&
  \seq(\eeq , \Veq) + \mbox{\rm [2nd order terms of $q^{\mu}$, $\Pi$ and $\sv\,^{\mu\nu}$]}
\\
 \Tne(\ene , \Vne , q^{\mu} , \Pi , \sv\,^{\mu\nu}) &=&
  \Teq(\eeq , \Veq) + \mbox{\rm [2nd order terms of $q^{\mu}$, $\Pi$ and $\sv\,^{\mu\nu}$]} \,,
\eqae
where we choose $\ene$, $\Vne$, $q^{\mu}$, $\Pi$ and $\sv\,^{\mu\nu}$ as the independent state variables of non-equilibrium state of each fluid element, and suffix ``ne'' denotes ``non-equilibrium''. 
The quantities $\ene$ and $\Vne$ are respectively the non-equilibrium specific internal energy~\footnote{
``Specific'' means the quantity per unit rest mass of fluid element.
} and specific volume, which are traditional variables. 
The others $q^{\mu}$, $\Pi$ and $\sv\,^{\mu\nu}$ are respectively the heat flux, bulk viscosity and shear viscosity, which are dissipative fluxes. 
In the above second order approximation, the quantities with suffix {\rm ``eq''} are the state variables of ``fiducial equilibrium state'',
\eqab
 \eeq &=&
 \text{internal energy of the fluid element in a fiducial equilibrium state}
 \\
 \Veq &=&
 \text{volume per unit rest mass of the fluid element in a fiducial equilibrium state} \,,
\eqae
\seqe
where the fiducial equilibrium state is defined as,
\begin{description}
\item[Definition of the fiducial equilibrium state:]
The fiducial equilibrium is the equilibrium state of fluid element of an imaginary perfect fluid (non-dissipative fluid) which possesses the same value of fluid velocity and rest mass density with our actual dissipative fluid. 
\end{description}
Then, $\seq(\eeq , \Veq)$ and $\Teq(\eeq , \Veq)$ in Eq.\eref{eq:eit.eos} are given by the equations of state for fiducial equilibrium state. 
Note that, under the second order approximation~\eref{eq:eit.eos}, the ``traditional variables'' are reduced to the state variables of fiducial equilibrium state. 
Also note that, because of Eq.\eref{eq:eit.ass3}, no first order term of independent dissipative flux appears in Eq.\eref{eq:eit.eos}. 
In summary, Eq.\eref{eq:eit.eos} is the expansion of non-equilibrium equations of state around the fiducial equilibrium state up to the second order of independent dissipative fluxes.
Thus, the dissipative fluxes under this condition are regarded as a non-equilibrium thermodynamic perturbation on the fiducial equilibrium state.~$\clubsuit$
\end{con}

As will be shown in Sec.\ref{sec:eqs.derivation}, the basic equations of EIT is derived using not only the assumptions~1 $\sim$~4 but also the supplemental condition~1. 
Then, one may regard the supplemental condition~1 as one of basic assumptions of EIT. 
However, in the study on non-equilibrium physics, there seem to be some efforts to go beyond the second order approximation required in supplemental condition~1~\cite{ref:eit_1,ref:eit_2}. 
Thus, in this manuscript, let us understand that the supplemental condition~1 is not a basic assumption but a supplemental condition to make the four basic assumptions work well.

A quantitative estimate of the strength of dissipative fluxes for a particular situation has been examined by Hiscock and Lindblom~\cite{ref:hl.applicable}. 
They investigated an ultra-relativistic gas including only a heat flux under the planar symmetry. 
We can recognize from the Hiscock-Lindblom's analysis that, for the system they investigated, the second order approximation of equations of state such as Eq.\eref{eq:eit.eos} is valid for the heat flux, $q^{\mu}$, satisfying the inequality,
\eqab
\label{eq:eit.validity}
 \dfrac{q}{\rhoeq\,\eeq} \lesssim 0.08898 \,,
\eqae
where $q \defeq \sqrt{q^{\mu}q_{\mu}}$. 
Note that the density of internal energy of fiducial equilibrium state, $\rhoeq\,\eeq$, includes the rest mass energy of the fluid. 
Therefore, the inequality~\eref{eq:eit.validity} implies that the supplemental condition~1 is appropriate when the heat flux is less than a few percent of the internal energy density including mass energy.

%%%%%%%%%%%%%%%%%%%%%%%%%%%%%%%%%%%%%%%%%%%%%%%%%%%%%%%%%%%%%%%%%%%%%%%%%%%%%%%%%%%%%%%%%%%%%%%%%%%%
\section{Basic quantities and basic equations of EIT}
\label{sec:eqs}

%%%%%%%%%%%%%%%%%%%%
\subsection{Meanings of basic quantities and equations}
\label{sec:eqs.meaning}

The basic quantities of dissipative fluid in the framework of EIT are,
\eqabnon
\begin{array}{lcl}
 u^{\mu}(x) &:&
  \mbox{velocity field of dissipative fluid
        (four-velocity of dissipative fluid element)}
\\
 \rhone(x)  &:&
  \mbox{rest mass density for non-equilibrium state}
\\
 \ene(x)    &:&
  \mbox{non-equilibrium \emph{specific} internal energy
        (internal energy \emph{per unit rest mass})}
\\
 \pne(x)    &:& \mbox{non-equilibrium pressure}
\\
 \Tne(x)    &:& \mbox{non-equilibrium temperature}
\\
 q^{\mu}(x) &:& \mbox{heat flux vector}
\\
 \Pi(x)     &:& \mbox{bulk viscosity}
\\
 \sv\,^{\mu\nu}(x) &:& \mbox{shear viscosity tensor}
\\
 g_{\mu\nu}(x) &:& \mbox{spacetime metric tensor \,.}
\end{array}
\eqaenon
The specific volume, $\Vne$ (volume per unit rest mass), which is one of traditional variables, is defined by
\eqab
 \Vne(x) \defeq \rhone(x)^{-1} \, .
\eqae
These quantities will appear in the basic equations of EIT.~\footnote{
In this manuscript, following the textbook of EIT~\cite{ref:eit_1,ref:eit_2}, we use the \emph{specific} scalar quantities, which are defined \emph{per unit rest mass}. 
On the other hand, some references of EIT~\cite{ref:hl.causality,ref:hl.applicable,ref:is_1,ref:is_2} use the \emph{density} of those scalar quantities, which are defined \emph{per unit three-volume} perpendicular to $u^{\mu}$. 
} 
All of the above quantities are the ``field'' quantities defined on spacetime manifold, and $x$ denotes the coordinate variables on spacetime. 
Quantities $\rhone$, $\ene$, $\pne$ and $\Tne$ are traditional variables, and quantities $q^{\mu}$, $\Pi$ and $\sv\,^{\mu\nu}$ are the dissipative fluxes (see assumption~2).

Note that, it is possible to determine the values of~$u^{\mu}$ and~$\rhone$ without referring to the other thermodynamic state variables. 
The fluid velocity, $u^{\mu}$, is simply defined as the average four-velocity of constituent particles in a fluid element. 
This definition of fluid velocity is called the N-frame by Israel~\cite{ref:is_1}. 
The rest mass density, $\rhone$, is simply defined by the rest mass per unit three-volume perpendicular to $u^{\mu}$. 
Because $u^{\mu}$ and $\rhone$ are determined without the knowledge of local non-equilibrium state, the fiducial equilibrium state is defined with referring to $u^{\mu}$ and $\rhone$ as explained in the supplemental condition~1.

Then, the rest mass current vector, $J^{\mu}$, is defined as
\eqb
\label{eq:eit.J}
 J^{\mu} \defeq \rhone\,u^{\mu} \,.
\eqe
The relations between the basic quantities and energy-momentum tensor, $T^{\mu\nu}$, of dissipative fluid are
\seqb
\label{eq:eit.quantity}
\eqab
 \rhone\,\ene &=& u_{\alpha}\,u_{\beta}\,T^{\alpha\beta}
\\
 q^{\mu}      &=& - \Delta^{\mu\alpha}\,u^{\beta}\,T_{\alpha\beta}
\\
\label{eq:eit.quantity.pPi}
 \pne + \Pi   &=& \dfrac{1}{3}\,\Delta_{\alpha\beta}\,T^{\alpha\beta}
\\
\label{eq:eit.quantity.sv}
 \sv\,^{\mu\nu} &=&
  \left[\,\Delta^{\mu\alpha}\,\Delta^{\nu\beta}
        - \dfrac{1}{3}\,\Delta^{\mu\nu}\,\Delta^{\alpha\beta} \,\right]\,T_{\alpha\beta} \,,
\eqae
\seqe
where $\Delta^{\mu\nu}$ is a projection tensor on perpendicular direction to $u^{\mu}$ defined as
\eqb
\label{eq:eit.Delta}
 \Delta^{\mu\nu} \defeq u^{\mu}\,u^{\nu} + g^{\mu\nu} \,.
\eqe
These relations~\eref{eq:eit.quantity} are simply the mathematically general decomposition of symmetric tensor $T^{\mu\nu}$. 
For the observer comoving with the fluid, $\rhone \ene$ is the temporal-temporal component of $T^{\mu\nu}$, $q^{\mu}$ the temporal-spatial component of $T^{\mu\nu}$, $\pne + \Pi$ the trace part of spatial-spatial component of $T^{\mu\nu}$, and $\sv\,^{\mu\nu}$ the trace-less part of spatial-spatial component of $T^{\mu\nu}$. 
Here, note that $\pne$ and $\Pi$ can not be distinguished by solely the relation~\eref{eq:eit.quantity.pPi}. 
However, we can distinguish $\pne$ and $\Pi$, when the equations of state are specified in which the pressure and bulk viscosity play different roles. 
Furthermore, as explained below, the basic equations of EIT are formulated so that $\pne$ and $\Pi$ are distinguished and obey different evolution equations. 
Thus, we find that, given the basic quantities of dissipative fluid, the energy-momentum tensor can be defined as
\eqb
\label{eq:eit.T}
 T^{\mu\nu} \defeq
  \rhone\,\ene\,u^{\mu}\,u^{\nu} + 2\,u^{(\mu} q^{\nu)}
  + (\pne + \Pi)\,\Delta^{\mu\nu} + \sv\,^{\mu\nu} \,,
\eqe
where the symmetrization $u^{(\mu} q^{\nu)}$ is defined as
\eqb
 u^{(\mu} q^{\nu)} \defeq
  \dfrac{1}{2}\,\left(\, u^{\mu}\,q^{\nu} + u^{\nu}\,q^{\mu} \,\right) \,.
\eqe

From the normalization of $u^{\mu}$, symmetry $T^{\mu\nu} = T^{\nu\mu}$ and relations~\eref{eq:eit.quantity}, we find some constraints on basic quantities of dissipative fluid~\cite{ref:is_1,ref:is_2}:
\seqb
\label{eq:eit.constraint}
\eqab
\label{eq:eit.uu}
 u^{\mu}u_{\mu} &=& -1
\\
\label{eq:eit.uq}
 u^{\mu}q_{\mu} &=& 0
\\
\label{eq:eit.sv_sym}
 \sv\,^{\mu\nu} &=& \sv\,^{\nu\mu}
\\
\label{eq:eit.usv}
 u_{\nu}\,\sv\,^{\mu\nu} &=& 0
\\
\label{eq:eit.sv_trace}
 \sv\,^{\mu}_{\,\,\,\mu} &=& 0 \,.
\eqae
\seqe
Of course, the metric is symmetric $g_{\mu\nu} = g_{\nu\mu}$. 
These constraints denote that the independent quantities are ten components of $g_{\mu\nu}$, three components of $u^{\mu}$, three components of $q^{\mu}$, five components of $\sv\,^{\mu\nu}$, five scalars $\rhone$, $\ene$, $\pne$, $\Tne$ and $\Pi$. 
Here recall that, according to assumption~3, some non-equilibrium equations of state should exist in order to guarantee the number of independent state variables. 
Such equations of state may be understood as constraints on state variables.

The ten components of metric $g_{\mu\nu}$ are determined by the Einstein equation,
\eqab
\label{eq:eit.einstein}
 G_{\mu\nu} = 8 \pi\, T_{\mu\nu} \,,
\eqae
where $G_{\mu\nu} \defeq R_{\mu\nu} - (1/2)\,R^{\alpha}_{\,\,\,\alpha}\,g_{\mu\nu}$ is the Einstein tensor, and $R_{\mu\nu}$ is the Ricci curvature tensor. 
Hence, in the framework of EIT, we need evolution equations to determine the remaining independent sixteen quantities $u^{\mu}$, $q^{\mu}$, $\sv\,^{\mu\nu}$, $\rhone$, $\ene$, $\pne$, $\Tne$ and $\Pi$.

Next, let us summarize the sixteen basic equations of EIT other than the Einstein equation. 
Hereafter, for simplicity, we omit the suffix ``eq'' of the state variables of fiducial equilibrium state,
\eqab
\label{eq:eit.suffix}
 \rho \defeq \rhoeq \quad,\quad
 \varepsilon \defeq \eeq \quad,\quad
 p \defeq \peq \quad,\quad
 T \defeq \Teq \quad,\quad
 s \defeq \seq \,.
\eqae
The five of the desired sixteen equations of EIT are given by the conservation law of rest mass, $J^{\mu}_{\,\,\,;\mu} = 0$, and that of energy-momentum, $T^{\mu\nu}_{\quad;\nu} = 0$ :
\seqb
\label{eq:eit.conservation}
\eqab
\label{eq:eit.mass}
 \ld{\rho} + \rho\,u^{\mu}_{\,\,\,;\mu}
  &=& 0
 \\
\label{eq:eit.energy}
 \rho\,\Bigl[\, \ld{\varepsilon} + (p+\Pi)\,\ld{V} \,\Bigr]
  &=&
  - q^{\mu}_{\,\,\,; \mu} - q^{\mu}\,\ld{u}_{\mu}
  - \sv\,^{\mu\nu}\,u_{\mu\,;\,\nu}
\\
\label{eq:eit.eom}
 \left(\, \rho\,\varepsilon + p + \Pi \,\right)\,\ld{u}\,^{\mu}
  &=&
  - \ld{q}\,^{\mu} + q_{\alpha}\,\ld{u}\,^{\alpha}\,u^{\mu} - u^{\alpha}_{\,\,\,; \alpha}\,q^{\mu}
  - q^{\alpha}\,u^{\mu}_{\,\,\,; \alpha}
  - \Bigl[\, (p+\Pi)_{, \alpha} + \sv\,^{\,\,\,\beta}_{\alpha\,\,\,; \beta} \,\Bigr]\,
    \Delta^{\alpha\mu} \,,
\eqae
\seqe
where the traditional variables are reduced to the state variables of fiducial equilibrium state due to the supplemental condition~1, and Eqs.\eref{eq:eit.conservation} retain only the first order dissipative corrections to the evolution equations of perfect fluid. 
Here, $\ld{Q}$ is the Lagrange derivative of quantity $Q$ defined by
\eqab
 \ld{Q} \defeq u^{\mu}\,Q_{;\mu} \,.
\eqae
And, Eq.\eref{eq:eit.mass} is given by $J^{\mu}_{\,\,\,;\mu} = 0$ which is the continuity equation (mass conservation) , Eq.\eref{eq:eit.energy} is given by $u_{\mu}T^{\mu\nu}_{\quad;\nu} = 0$ which is the energy conservation and corresponds to the first law of non-equilibrium thermodynamics in the EIT, and Eq.\eref{eq:eit.eom} is given by $\Delta^{\mu\alpha}T^{\,\,\,\beta}_{\alpha\,\,\,;\beta} = 0$ which is the Euler equation (equation of motion of dissipative fluid).

Here, let us note the relativistic effects and number of independent equations. 
The relativistic effects are $q^{\mu}\,\ld{u}_{\mu}$ in Eq.\eref{eq:eit.energy}, and $(p+\Pi)\,\ld{u}\,^{\mu}$ and the terms including $q^{\mu}$ in Eq.\eref{eq:eit.eom}. 
Those terms do not appear in non-relativistic EIT~\cite{ref:eit_1,ref:eit_2}. 
And, due to the constraint of normalization~\eref{eq:eit.uu}, three components of Euler equation~\eref{eq:eit.eom} are independent, and one component is dependent. 
Totally, the five equations are independent in the set of equations~\eref{eq:eit.conservation}.

The nine of desired sixteen equations of EIT are the evolution equations of dissipative fluxes, whose derivation are reviewed in next subsection using the assumptions~1 $\sim$~4 and supplemental condition~1. 
According the next subsection or references of EIT~\cite{ref:hl.causality,ref:eit_1,ref:eit_2,ref:is_1}, the evolution equations of dissipations are
\seqb
\label{eq:eit.dissipation}
\eqab
\label{eq:eit.heat}
 \tauh\,\ld{q}\,^{\mu} &=&
  - \left[ 1 + \lambda\,T^2
               \left(\dfrac{\tauh}{2 \lambda T^2}\,u^{\nu} \right)_{; \nu} \right] q^{\mu}
  - \lambda\,T\,\ld{u}\,^{\mu} + \tauh ( q^{\nu} \ld{u}_{\nu} )\,u^{\mu}
\\
\nonumber
 &&
  - \lambda\,\Delta^{\mu\nu}\,\left[\, T_{,\nu} - T^2\,\left\{
        \inthb\,\Pi_{,\nu} + (1-\ghb)\,\Pi\,\beta_{{\rm bh}\,,\nu}
      + \inths\,\sv\,^{\,\,\,\alpha}_{\nu\,\,\,;\alpha}
      + (1-\ghs)\,\beta_{{\rm hs}\,,\alpha}\,\sv\,^{\alpha}_{\,\,\,\nu}
    \right\}\,\,\right]
\\
\label{eq:eit.bulk}
 \taub\,\ld{\Pi} &=&
  - \left[ 1 + \zeta\,T\,
               \left(\dfrac{\taub}{2 \zeta T}\,u^{\mu} \right)_{; \mu} \right] \Pi
  - \zeta\,u^{\mu}_{\,\,\,;\mu}
  + \zeta\,T\,\left(\, \inthb\, q^{\mu}_{\,\,\,;\mu}
                     + \ghb\,q^{\mu}\,\beta_{{\rm hb}\,,\mu} \,\right)
\\
\label{eq:eit.shear}
 \taus\,\bigl(\,\sv\,^{\mu\nu}\bigr)^{\bullet} &=&
  - \left[ 1 + 2\,\eta\,T\,
               \left(\dfrac{\taus}{4 \eta T}\,u^{\alpha} \right)_{; \alpha} \right]
    \sv\,^{\mu\nu}
  + 2\,\taus\,\ld{u}_{\alpha}\,\sv\,^{\alpha\,(\mu}\,u^{\nu)}
\\
\nonumber
 &&
  - 2\,\eta\,
  \tssp{
    u^{\mu;\nu} - T\,\left\{\inths\,q^{\mu;\nu} + \ghs\,\inths^{\,\,\,\,,\mu}\,q^{\nu}\right\}
    } \,,
\eqae
\seqe
where the symbolic operation $\tssp{A^{\mu\nu}}$ in the last term in Eq.\eref{eq:eit.shear} denotes the traceless symmetrization of a tensor $A^{\mu\nu}$ in the perpendicular direction to $u^{\mu}$,
\eqab
\label{eq:eit.sst}
\tssp{A^{\mu\nu}} \defeq
 \Delta^{\mu\alpha}\,\Delta^{\nu\beta}\,A_{(\alpha\beta)}
 - \dfrac{1}{3} \,\Delta^{\mu\nu}\,\Delta^{\alpha\beta}A_{\alpha\beta} \,.
\eqae
We find $\sv\,^{\mu\nu} = \tssp{T^{\mu\nu}}$ by Eq.\eref{eq:eit.quantity.sv}, and $\tssp{\sv\,^{\mu\nu}} = \sv\,^{\mu\nu}$.

The meanings of coefficients appearing in Eq.\eref{eq:eit.dissipation} are,
\seqb
\label{eq:eit.coeff}
\eqb
\label{eq:eit.coeff.classic}
\begin{array}{lcl}
 \lambda &:& \mbox{heat conductivity}
\\
 \zeta   &:& \mbox{bulk viscous rate}
\\
 \eta    &:& \mbox{shear viscous rate}
\\
 \tauh &:& \mbox{relaxation time of heat flux $q^{\mu}$}
\\
 \taub &:& \mbox{relaxation time of bulk viscosity $\Pi$}
\\
 \taus &:& \mbox{relaxation time of shear viscosity $\sv\,^{\mu\nu}$ ,}
\end{array}
\eqe
and
\eqb
\label{eq:eit.coeff.eit}
\begin{array}{lcl}
 \inthb &:&
   \mbox{interaction coefficient between dissipative fluxes $q^{\mu}$ and $\Pi$}
\\
 \inths &:&
   \mbox{interaction coefficient between dissipative fluxes $q^{\mu}$ and $\sv\,^{\mu\nu}$}
\\
 \ghb &:&
   \mbox{interaction coefficient between thermodynamic forces of $q^{\mu}$ and $\Pi$}
\\
 \ghs &:&
  \mbox{interaction coefficient between thermodynamic forces of $q^{\mu}$ and $\sv\,^{\mu\nu}$ ,}
\end{array}
\eqe
\seqe
where the thermodynamic forces of $q^{\mu}$, $\Pi$ and $\sv\,^{\mu\nu}$, which we express respectively by symbols $X\ls{h}^{\mu}$, $X\ls{b}$ and $X\ls{s}^{\mu\nu}$, are the quantities appearing in the bilinear form~\eref{eq:eit.bilinear} as $\sigma\ls{s} = q_{\mu}\,X\ls{h}^{\mu} + \Pi\,X\ls{h} + \sv_{\mu\nu}\,X\ls{s}^{\mu\nu}$. 
Details are explained in Sec.\ref{sec:eqs.derivation}

In general, the above ten coefficients are functions of state variables of fiducial equilibrium state. 
Those functional forms should be determined by some micro-scopic theory, turbulence theory or experiment of dissipative fluxes, but it is out of the scope of this manuscript.

The coefficients in list~\eref{eq:eit.coeff.classic} are already known in the traditional laws of dissipations and Maxwell-Cattaneo laws summarized in appendix~\ref{app:mc}. 
Note that the existence of relaxation times of dissipative fluxes make the evolution equations~\eref{eq:eit.dissipation} retain the causality of dissipative phenomena. 
The relaxation time, $\tauh$, is the time scale in which a non-stationary heat flux relaxes to a stationary heat flux. 
The other relaxation times, $\taub$ and $\taus$, have the same meaning for viscosities. 
These are positive by definition, 
\eqab
\label{eq:eit.coeff.relax}
 \tauh > 0 \quad,\quad
 \taub > 0 \quad,\quad
 \taus > 0 \,.
\eqae
Concerning the transport coefficients, $\lambda$, $\zeta$ and $\eta$, the non-negativity of them is obtained by the requirement~(4-a) in assumption~4 as explained in next subsection, 
\eqab
\label{eq:eit.coeff.trans}
 \lambda \ge 0 \quad,\quad
 \zeta \ge 0 \quad,\quad
 \eta \ge 0 \,.
\eqae

The coefficients in list~\eref{eq:eit.coeff.eit} denotes that the EIT includes the interaction among dissipative fluxes, while the traditional laws of dissipations and Maxwell-Cattaneo laws do not. 
(See appendix~\ref{app:mc} for a short summary.) 
Concerning the interaction among dissipative fluxes, Israel~\cite{ref:is_1} has introduced an approximation into the evolution equations~\eref{eq:eit.dissipation}. 
Israel ignores the gradients of fiducial equilibrium state variables, as summarized in the end of next subsection. 
Then, Eq.\eref{eq:eit.dissipation} can be simplified by discarding terms including the gradients. 
Those simplified equations are shown in Eq.\eref{eq:eqs.derivation.dissipation_israel}.

Given the meanings of all quantities which appear in Eq.\eref{eq:eit.dissipation}, we can recognize a thermodynamical feature of Eq.\eref{eq:eit.dissipation}. 
Recall that the dissipative phenomena are thermodynamically \emph{irreversible processes}. 
Then, reflecting the irreversible nature, the evolution equations of dissipative fluxes~\eref{eq:eit.dissipation} are not time-reversal invariant, i.e. Eq.~\eref{eq:eit.dissipation} are not invariant under the replacement, $u^{\mu} \to - u^{\mu}$ and $q^{\mu} \to -q^{\mu}$.

Here, let us note the relativistic effects and number of independent evolution equations of dissipative fluxes. 
The relativistic effects are three terms including $\ld{u}\,^{\mu}$ and three terms of the form $(\,\bigcirc\,u^{\mu})_{;\mu}$ in right-hand sides of Eq.\eref{eq:eit.dissipation}. 
Those terms disappear in non-relativistic EIT~\cite{ref:eit_1,ref:eit_2}. 
And, due to the constraints in Eqs.\eref{eq:eit.constraint} except Eq.\eref{eq:eit.uu}, the three components of evolution equation~\eref{eq:eit.heat} and five components of evolution equation~\eref{eq:eit.shear} are independent. 
Totally, nine equations are independent in the set of equations~\eref{eq:eit.dissipation}.

From the above, we have fourteen independent evolution equations in Eqs.\eref{eq:eit.conservation} and~\eref{eq:eit.dissipation}. 
We need the other two equations to determine the sixteen quantities which appear in Eqs.\eref{eq:eit.conservation} and~\eref{eq:eit.dissipation}. 
Those two equations, under the supplemental condition~1, are the equations of state of fiducial equilibrium state. 
They are expressed, for example, as
\eqab
\label{eq:eit.eos.eq}
 p = p(\varepsilon , V) \quad,\quad
 T = T(\varepsilon , V) \,.
\eqae
The concrete forms of Eq.\eref{eq:eit.eos.eq} can not be specified unless the dissipative matter composing the fluid is specified.

In summary, the basic equations of EIT, under the supplemental condition~1, are Eqs.\eref{eq:eit.conservation}, \eref{eq:eit.dissipation} and~\eref{eq:eit.eos.eq} with constraints~\eref{eq:eit.constraint}, and furthermore the Einstein equation~\eref{eq:eit.einstein} for the evolution of spacetime metric. 
With those basic equations, it has already been known that the causality is retained for dissipative fluids which are thermodynamically stable. 
Here the ``thermodynamic stability'' means that, for example, the heat capacity and isothermal compressibility are positive~\cite{ref:hl.causality}. 
The positive heat capacity and positive isothermal compressibility are the very usual and normal property of real materials. 
We recognize that \emph{the EIT is a causal hydrodynamics for dissipative fluids made of ordinary matters}.

%%%%%%%%%%%%%%%%%%%%
\subsection{Derivation of evolution equations of dissipations}
\label{sec:eqs.derivation}

Let us proceed to the derivation of Eqs.\eref{eq:eit.dissipation}. 
In order to obtain them, we refer to the assumption~4 and need the non-equilibrium entropy current vector, $S\ls{ne}^{\,\,\,\,\mu}$. 
The entropy current, $S\ls{ne}^{\,\,\,\,\mu}$, is a member of dissipative fluxes (see assumption~2). 
Hereafter, we choose $q^{\mu}$, $\Pi$ and $\sv\,^{\mu\nu}$ as the three independent dissipative fluxes (see assumption~3). 
Then $S\ls{ne}^{\,\,\,\,\mu}$ is a dependent state variable and should be expanded up to the second order of independent dissipative fluxes under the supplemental condition~1~\cite{ref:eit_1,ref:eit_2,ref:is_1},
\eqab
\label{eq:eqs.derivation.Sne}
 S\ls{ne}^{\,\,\,\,\mu} \defeq
 \rhone\,\sne\,u^{\mu} + \dfrac{1}{T}\,q^{\mu}
 + \inthb\,\Pi\,q^{\mu} + \inths\,q_{\nu}\,\sv\,^{\nu\mu} \, ,
\eqae
where we assume the \emph{isotropic} equations of state which will be explained below, the factor $\rhone\,\sne$ in the first term is expanded up to the second order of independent dissipative fluxes due to the supplemental condition~1, the second term $T^{-1} q^{\mu}$ is the first order term of heat flux due to the meaning of ``heat'' already known in the ordinary equilibrium thermodynamics, and the third and fourth terms express the interactions between heat flux and viscosities as noted in list~\eref{eq:eit.coeff.eit}. 
These interactions between heat and viscosities are one of significant properties of EIT, while traditional laws of dissipations and Maxwell-Cattaneo laws summarized in appendix~\ref{app:mc} do not include these interactions.

Before proceeding to the discussion on the bilinear form of entropy production rate, we should give two remarks on Eq.\eref{eq:eqs.derivation.Sne}: 
First remark is on the first order term of heat flux, $T^{-1}q^{\mu}$. 
One may think that this term is inconsistent with Eq.\eref{eq:eit.ass3}, since the differential, 
$\pdline{(T^{-1}q^{\mu})}{q^{\mu}} = T^{-1}$, does not vanish at local equilibrium limit. 
However, recall that the fluid velocity, $u^{\mu}$, depends on the dissipative fluxes. 
Then, we expect a relation, $\rhone\,\sne\,(\pdline{u^{\mu}}{q^{\mu}}) = - T^{-1}$, by which Eq.\eref{eq:eit.ass3} is satisfied. 
The evolution equations of EIT should yield $u^{\mu}$ so that $S\ls{ne}^{\,\,\,\,\mu}$ satisfies Eq.\eref{eq:eit.ass3}.

Second remark is on the second order terms of dissipative fluxes in Eq.\eref{eq:eqs.derivation.Sne}, which reflect the notion of isotropic equations of state. 
Considering a general form of those terms relates to considering a general form of non-equilibrium equations of state. 
In general, there may be a possibility for non-equilibrium state that equations of state depend on a special direction, e.g. a direction of spinor of constituent particles, a direction of defect of crystal structure in a solid or liquid crystal system, a direction originated from some turbulent structure, and so on, which reflect a rather micro-scopic structure of the system under consideration. 
If a dependence on such a special direction arises in non-equilibrium equations of state, then the entropy current, $S\ls{ne}^{\,\,\,\,\mu}$, may depend on some tensors reflecting the special direction, and its most general form up to the second order of independent dissipative fluxes is~\cite{ref:hl.causality,ref:is_2,ref:eit_1,ref:eit_2}
\eqb
\label{eq:eqs.derivation.Sne.general}
\begin{array}{rl}
 S\ls{ne}^{\,\,\,\,\mu} \defeq & \mbox{Eq.\eref{eq:eqs.derivation.Sne}}
 + \inthb'\,\Pi\,A^{\mu\alpha}q_{\alpha}
 + \inths'\,B_{\alpha\beta}q^{\alpha}\sv\,^{\beta\mu}
 + \inths''\,C^{\alpha\beta\gamma\mu}q_{\alpha}\sv_{\beta\gamma}
 + \inths'''\,D^{\alpha\beta}\sv_{\alpha\beta}q^{\mu} \\
 &
 + \beta'\ls{bs}\,\Pi\,\sv\,^{\mu\alpha}E_{\alpha}
 + \beta''\ls{hs}\Pi\,\sv_{\alpha\beta}F^{\alpha\beta\mu}
 + \beta'\ls{bb}\,\Pi^2\,G^{\mu}
 + \beta'\ls{hh}\,H^{\alpha}q_{\alpha}q^{\mu}
 + \beta''\ls{hh}\,I^{\alpha\beta\mu}q_{\alpha}q_{\beta} \\
 &
 + \beta'\ls{ss}\,J_{\alpha\beta\gamma}\sv\,^{\alpha\beta}\sv\,^{\gamma\mu}
 + \beta''\ls{ss}\,K^{\mu}\sv\,^{\alpha\beta}\sv_{\alpha\beta}
 \, ,
\end{array}
\eqe
where $A^{\mu\nu}\,\cdots\,K^{\mu}$ are the tensors reflecting the special direction.
Although Eq.\eref{eq:eqs.derivation.Sne.general} is the most general form of $S\ls{ne}^{\,\,\,\,\mu}$, the inclusion of such a special direction raises an inessential mathematical confusion in following discussions.~\footnote{
There may be a possibility that the factor tensors are gradients of fiducial equilibrium state variable, e.g. $K^{\mu} \propto \ene^{\,\,\,\,,\mu}$. 
However, in thermodynamics, it is naturally expected that equations of state do not depend on gradients of state variables but depend only on the state variables themselves. 
Further, if complete non-equilibrium equations of state do not include gradients of state variables, their Tayler expansion can not include gradients in the expansion factors. 
}
Furthermore, recall that, usually, such a special direction of \emph{micro}-scopic structure does not appear in ordinary equilibrium thermodynamics which describes the \emph{macro}-scopic properties of the system. 
Thus, under the supplemental condition~1 which restricts our attention to non-equilibrium states \emph{near} equilibrium states, it may be expected that such a special direction does not appear in non-equilibrium equations of state. 
Let us assume the \emph{isotropic equations of state} in which the directional dependence does not exist, and adopt Eq.\eref{eq:eqs.derivation.Sne} as the equation of state for $S\ls{ne}^{\,\,\,\,\mu}$.~\footnote{
When one specifies the material composing the dissipative fluid, and if its non-equilibrium equations of state have some directional dependence, then the same procedure given in Sec.~\ref{sec:eqs.derivation} provides the basic equations of EIT depending on a special direction.}

Here, since a non-equilibrium factor, $\rhone\, \sne$, appears in Eq.\eref{eq:eqs.derivation.Sne}, we need to show the non-equilibrium equation of state for it. 
Adopting the isotropic assumption, the non-equilibrium equation of state~\eref{eq:eit.eos.sne} for $\sne$ becomes~\cite{ref:eit_1,ref:eit_2},
\eqab
\label{eq:eit.eos.sne.isotropic}
 \rhone\,\sne(\ene , \Vne , q^{\mu} , \Pi , \sv\,^{\mu\nu}) =
 \rho\,s(\varepsilon , V) - a\ls{h}\,q^{\mu}\,q_{\mu}
 - a\ls{b}\,\Pi^2 - a\ls{s}\,\sv\,^{\mu\nu}\,\sv_{\mu\nu} \,,
\eqae
where the suffix ``eq'' of state variables of fiducial equilibrium state in right-hand side are omitted as noted in Eq.\eref{eq:eit.suffix}, the expansion coefficients, $a\ls{h}$, $a\ls{b}$ and $a\ls{s}$, are functions of fiducial equilibrium state variables, and the minus sign in front of them expresses that the non-equilibrium entropy is less than the fiducial equilibrium entropy. 
Concrete forms of those coefficients will be obtained below.

On the specific entropy of fiducial equilibrium state, $s(\varepsilon,V)$, the first law of thermodynamics for fiducial equilibrium state is important in calculating the bilinear form of entropy production rate~\eref{eq:eit.bilinear}, $T\ld{s} = \ld{\varepsilon} + p\,\ld{V}$. 
Combining the first law of fiducial equilibrium state with the energy conservation~\eref{eq:eit.energy}, we find
\eqb
\label{eq:eqs.1stlaw}
 \rho\,\ld{s} + \dfrac{1}{T}\,q^{\mu}_{\,\,\,; \mu} =
  - \dfrac{1}{T}\,
    \bigl(\,u^{\mu}_{\,\,\,; \mu}\,\Pi
          + \ld{u}_{\mu} q^{\mu} + u_{\mu;\nu}\,\sv\,^{\mu\nu} \,\bigr) \,,
\eqe
where Eq.\eref{eq:eit.mass} is used in deriving the first term in right-hand side. 
This relation is used in following calculations.

Given the above preparation, we can proceed to calculation of the bilinear form of entropy production rate. 
The entropy production rate is defined as the divergence, $\sigma\ls{s} \defeq S\ls{ne \,\,\,;\mu}^{\,\,\,\,\mu}$, as already given in assumption~4. 
Then, according to Eq.\eref{eq:eit.bilinear}, $\sigma\ls{s}$ should be rearranged to the bilinear form~\cite{ref:eit_1,ref:eit_2},
\eqb
\label{eq:eqs.eos.sigmas}
 \sigma\ls{s} \defeq S\ls{ne \,\,\,;\mu}^{\,\,\,\,\mu}
 = q_{\mu}\,X^{\mu}\ls{h} + \Pi\,X\ls{b} + \sv_{\mu\nu}\,X^{\mu\nu}\ls{s} \,,
\eqe
where the factors, $X^{\mu}\ls{h}$, $X\ls{b}$ and $X^{\mu\nu}\ls{s}$, are the thermodynamic forces. 
To determine the concrete forms of thermodynamic forces, let us carry out the calculation of the divergence, $S\ls{ne \,\,\,;\mu}^{\,\,\,\,\mu}$ of Eq.\eref{eq:eqs.derivation.Sne}. 
We find immediately that the divergence includes the differentials of $\inthb$ and $\inths$ as, $S\ls{ne \,\,\,;\mu}^{\,\,\,\,\mu} = \Pi\,q^{\mu}{\inthb}_{,\mu} + \sv\,^{\mu\nu}q_{\nu}{\inths}_{,\mu} + \cdots$\,. 
The assumptions~1 $\sim$~4 can not determine whether the term $\Pi\,q^{\mu}{\inthb}_{,\mu}$ should be put into $q_{\mu}\,X^{\mu}\ls{h}$ or $\Pi\,X\ls{b}$ in Eq.\eref{eq:eqs.eos.sigmas}, and whether the term $\sv\,^{\mu\nu}q_{\nu}{\inths}_{,\mu}$ should be put into $q_{\mu}\,X^{\mu}\ls{h}$ or $\sv_{\mu\nu}\,X^{\mu\nu}\ls{s}$ in Eq.\eref{eq:eqs.eos.sigmas}. 
Hence, we introduce additional factors, $\ghb$ and $\ghs$, to divide those terms so that the three terms in Eq.\eref{eq:eqs.eos.sigmas} become~\cite{ref:hl.causality}
\eqb
 \begin{array}{rcl}
 q_{\mu}\,X^{\mu}\ls{h} &=&
  q_{\mu}\,\Bigl[\,(1-\ghb)\, \Pi\, \inthb^{\,\,\,\,,\mu}
                 + (1-\ghs)\, \sv\,^{\mu\nu} {\inths}_{,\nu} + \cdots \,\Bigr] \\
 \Pi\,X\ls{b} &=&
  \Pi\,\Bigl[\,\ghb\,q^{\mu} {\inthb}_{,\mu} + \cdots \,\Bigr] \\
 \sv_{\mu\nu}\,X^{\mu\nu}\ls{s} &=&
  \sv_{\mu\nu}\,\Bigl[\, \ghs\,q^{\nu}\inths^{\,\,\,\,,\mu} + \cdots \,\Bigr] \,.
 \end{array}
\eqe
Note that $\ghb$ and $\ghs$ are included in thermodynamic forces. 
The factor $\ghb$ connects $X\ls{h}^{\mu}$ and $X\ls{b}$, and $\ghs$ connects $X\ls{h}^{\mu}$ and $X\ls{s}^{\mu\nu}$. 
Therefore, we can understand that these factors, $\ghb$ and $\ghs$, are the kind of interaction coefficients among thermodynamic forces as noted in list~\eref{eq:eit.coeff.eit}.

Then, using Eqs.\eref{eq:eit.eos.sne.isotropic} and~\eref{eq:eqs.1stlaw}, we obtain the concrete forms of $X$'s,
\eqb
\label{eq:eqs.force}
 \begin{array}{rcl}
  X\ls{h}^{\mu} &=&
   - 2\,a\ls{h}\,\ld{q}\,^{\mu} - \bigl(a\ls{h} u^{\alpha} \bigr)_{; \alpha}\,u^{\mu}
   - \dfrac{1}{T}\,\ld{u}\,^{\mu} - \dfrac{1}{T^2}\,T^{, \mu} \\
 &&
   + \inthb\,\Pi^{, \mu} + (1-\ghb)\,\inthb^{\,\,\,\,, \mu}\,\Pi
   + \inths\,\sv\,^{\mu\alpha}_{\quad; \alpha}
   + (1-\ghs)\,\inths\,_{\!,\alpha}\,\sv\,^{\alpha\mu} \\[2mm]
  X\ls{b} &=&
   - 2\,a\ls{b}\,\ld{\Pi} - \bigl(a\ls{b} u^{\alpha} \bigr)_{; \alpha}\,\Pi
   - \dfrac{1}{T}\,u^{\mu}_{\,\,\,; \mu}
   + \inthb\,q^{\mu}_{\,\,\,; \mu} + \ghb\,\inthb\,_{\!,\mu}\,q^{\mu} \\[2mm]
  X\ls{s}^{\mu\nu} &=&
   - 2\,a\ls{s}\,\bigl(\sv\,^{\mu\nu}\bigr)^{\bullet}
   - \bigl(a\ls{s} u^{\alpha} \bigr)_{; \alpha}\,\sv\,^{\mu\nu}
   - \dfrac{1}{T}\,u^{\mu ; \nu}
   + \inths\,q^{\mu ; \nu} + \ghb\,\inthb^{\,\,\,\,,\mu}\,q^{\nu} \,.
 \end{array}
\eqe
Obviously these thermodynamic forces include $\ld{q}\,^{\mu}$, $\ld{\Pi}$ and $(\sv\,^{\mu\nu})^{\bullet}$. 
Then, as reviewed bellow, making use of this fact and assumption~4 enables us to obtain the evolution equations of dissipative fluxes in the form, $[\mbox{dissipative flux}]^{\bullet} = \cdots$.

Thermodynamic forces, $X\ls{h}^{\mu}$ and $X\ls{s}^{\mu\nu}$ shown in Eq.\eref{eq:eqs.force}, have some redundant parts. 
For $X\ls{h}^{\mu}$, its component parallel to $u^{\mu}$ is redundant, because we find $q_{\mu}\,X\ls{h}^{\mu} = q^{\mu}\,(\Delta_{\mu\nu}X\ls{h}^{\nu})$ due to the constraint~\eref{eq:eit.uq}.
For $X\ls{s}^{\mu\nu}$, its trace part and components parallel to $u^{\mu}$ are redundant, because we find $\sv_{\mu\nu}\,X^{\mu\nu}\ls{s} = \sv_{\mu\nu}\,\tssp{X^{\mu\nu}\ls{s}}$ due to the relation $\sv\,^{\mu\nu} = \tssp{\sv\,^{\mu\nu}}$, where the operation $\tssp{\,\,\cdot\,\,}$ are defined in Eq.\eref{eq:eit.sst}. 
Therefore, Eq.\eref{eq:eqs.eos.sigmas} becomes
\eqb
\label{eq:eqs.eos.sigmas_2}
 \sigma\ls{s} \defeq S\ls{ne \,\,\,;\mu}^{\,\,\,\,\mu}
 = q^{\mu}\,(\Delta_{\mu\nu}X\ls{h}^{\nu}) + \Pi\,X\ls{b}
   + \sv_{\mu\nu}\,\tssp{X^{\mu\nu}\ls{s}} \,.
\eqe
This is understood as an equation of state for $\sigma\ls{s}$. 
Hence, we obtain the following relations due to supplemental condition~1 and Eq.\eref{eq:eit.ass3},
\eqab
\label{eq:eqs.X}
 \Delta_{\mu\nu}X\ls{h}^{\nu} = b\ls{h}\,q_{\mu} \quad,\quad
 X\ls{b} = b\ls{b}\,\Pi \quad,\quad
 \tssp{X\ls{s}^{\mu\nu}} = b\ls{s}\,\sv\,^{\mu\nu} \,,
\eqae
where the coefficients, $b\ls{h}$, $b\ls{b}$ and $b\ls{s}$, are functions of fiducial equilibrium state variables. 
Concrete forms of them are determined as follows: 
According to the requirement (4-b) in assumption~4, Eq.\eref{eq:eqs.X} should be consistent with existing phenomenologies even in non-relativistic cases. 
As such reference phenomenologies, we refer to the \emph{Maxwell-Cattaneo laws}, which are summarized in appendix~\ref{app:mc}. 
By comparing Eq.\eref{eq:eqs.X} with the Maxwell-Cattaneo laws in Eq.\eref{eq:mc.mc}, the unknown coefficients are determined~\cite{ref:eit_1,ref:eit_2},
\eqab
\label{eq:eqs.coeff}
 a\ls{h} = \dfrac{\tauh}{2\,\lambda\,T^2} \quad,\quad
 a\ls{b} = \dfrac{\taub}{2\,\zeta\,T} \quad,\quad
 a\ls{s} = \dfrac{\taus}{4\,\eta\,T} \quad,\quad
 b\ls{h} = \dfrac{1}{\lambda\,T^2} \quad,\quad
 b\ls{b} = \dfrac{1}{\zeta\,T} \quad,\quad
 b\ls{s} = \dfrac{1}{2\,\eta\,T} \,,
\eqae
where $\lambda$, $\zeta$, $\eta$, $\tauh$, $\taub$ and $\taus$ are shown in list~\eref{eq:eit.coeff.classic}. 
By Eq.\eref{eq:eit.bilinear}, non-negativity of coefficients~\eref{eq:eit.coeff.trans} is obtained.

Then, by substituting those coefficients~\eref{eq:eqs.coeff} into the concrete forms of thermodynamic forces given in Eq.\eref{eq:eqs.force}, Eq.\eref{eq:eqs.X} are rearranged to the form of evolution equations, $\tau\ls{h}\,\ld{q}\,^{\mu} = \cdots$ , $\tau\ls{b}\,\ld{\Pi} = \cdots$ and $\tau\ls{s}\,(\sv\,^{\mu\nu})^{\bullet} = \cdots$~\cite{ref:hl.causality,ref:eit_1,ref:eit_2}.
These are the evolution equations of dissipative fluxes shown in Eq.\eref{eq:eit.dissipation}.

Finally in this section, summarize a discussion given in an original work of EIT~\cite{ref:is_1}: 
Under the supplemental condition~1, the dissipative fluxes appearing in Eqs.\eref{eq:eit.conservation} and~\eref{eq:eit.dissipation} are not so strong. 
Then, there may be many actual situations that the gradients of fiducial equilibrium state variables are also week. 
Motivated by this consideration, Israel~\cite{ref:is_1} has introduced an additional supplemental condition:
\begin{con}[A strong restriction by Israel]
The order of gradient of any state variables of fiducial equilibrium state is at most the same order with dissipative fluxes,
\eqab
 k\,\pd{[\mbox{\rm fiducial equilibrium state variables}]}{x^{\mu}}
 \lesssim O([\mbox{\rm dissipative fluxes}]) \,,
\eqae
where $k$ is an appropriate numerical factor to make the left- and right-hand sides have the same dimension.~$\clubsuit$
\end{con}
This condition restricts the applicable range of EIT narrower than the supplemental condition~1. 
However, as discussed by Israel~\cite{ref:is_1}, if one adopts this condition, then the evolution equations of dissipative fluxes~\eref{eq:eit.dissipation} are simplified by discarding the terms of [dissipative fluxes]$\times$[gradients of fiducial equilibrium state variable],
\seqb
\label{eq:eqs.derivation.dissipation_israel}
\eqab
 \tauh\,\ld{q}\,^{\mu}
 &=&
  - q^{\mu} - \lambda T\,\ld{u}\,^{\mu} - \lambda\,\Delta^{\mu\nu}\,\Bigl[\, 
     T_{,\nu} - T^2\,\Bigl(\,\inthb\,\Pi_{,\nu}
                           + \inths\,\sv\,^{\,\,\,\alpha}_{\nu\,\,\,\,;\alpha} \,\Bigr)
    \Bigr] \\
 \taub\,\ld{\Pi}
 &=&
  - \Pi - \zeta\,u^{\mu}_{\,\,\,;\mu} + \beta\ls{hb}\,\zeta\,T\,q^{\mu}_{\,\,\,;\mu} \\
 \taus\,\bigl(\,\sv\,^{\mu\nu}\bigr)^{\bullet}
 &=&
  - \sv\,^{\mu\nu}
  - 2\,\eta\,\tssp{u^{\mu;\nu} - T\,\inths\,q^{\mu;\nu}} \,.
\eqae
\seqe
However, Hiscock and Lindblom~\cite{ref:hl.causality} point out that the condition~2 may not necessarily be acceptable, for example, for the stellar structure in which the gradients of temperature and pressure play the important role. 
Furthermore, as implied by Eq.\eref{eq:eit.dissipation}, when the interaction coefficients among thermodynamic forces, $\ghb$ and $\ghs$, are very large, the terms including differentials ${\inthb}_{,\mu}$ and ${\inths}_{,\mu}$ can not necessarily be ignored.

%%%%%%%%%%%%%%%%%%%%%%%%%%%%%%%%%%%%%%%%%%%%%%%%%%%%%%%%%%%%%%%%%%%%%%%%%%%%%%%%%%%%%%%%%%%%%%%%%%%%
\section{EIT and Radiative Transfer}
\label{sec:rad}

%%%%%%%%%%%%%%%%%%%%
\subsection{Overview of one Limit of EIT}

As mentioned at the end of Sec.\ref{sec:eqs.meaning}, if and only if the dissipative fluid is made of thermodynamically normal matter with positive heat capacity and positive isothermal compressibility (the ``ordinary matter''), then the EIT is a causally consistent phenomenology of the dissipative fluid with including interactions among dissipations~\cite{ref:hl.causality}. 
Then, it is necessary to make a remark on the \emph{hydrodynamic and/or thermodynamic treatment of non-equilibrium radiation field}, because, as will be explained below, a radiation field changes its character according to the situation in which the radiation field is involved. 
Here the ``radiation field'' means the matters composed of non-self-interacting particles such as gravitons, neutrinos (if it is massless) and photons (with neglecting the quantum electrodynamical pair creation and annihilation of photons in very high temperature states). 
Hereafter, the ``photon''means the constituent particle of radiation field.

Some special properties of non-equilibrium state of radiation field have been investigated: 
Wildt~\cite{ref:noneq.rad_1} found some strange property of entropy production process in the radiation field, and Essex~\cite{ref:noneq.rad_2,ref:noneq.rad_3} recognized that the bilinear form of $\sigma\ls{s}$ given in Eq.\eref{eq:eit.bilinear} is incompatible with the non-equilibrium state of radiation field in \emph{optically thin matters}. 
This denotes that the EIT can not be applied to non-equilibrium radiation fields in optically thin matters. 
In other words, the formalism of EIT becomes applicable to non-equilibrium radiative transfer at the limit of \emph{vanishing mean-free-path of photons} as considered by Udey and Israel~\cite{ref:rad.dense_2} and by Fort and Llebot~\cite{ref:rad.dense_1}. 
And no thermodynamic formulation of non-equilibrium radiation field in optically thin matters had not been constructed until some years ago. 
Then, one of present authors constructed explicitly a \emph{steady state} thermodynamics for a stationary non-equilibrium radiation field in optically thin matters~\cite{ref:noneq.rad_4}, where the energy flow in the non-equilibrium state is stationary. 
As shown in this section, the steady state thermodynamics for a radiation field, which is different from EIT, is inconsistent with the bilinear form of entropy production rate. 
Inconsistency of EIT with optically thin radiative transfer is not explicitly recognized in the standard references of EIT~\cite{ref:eit_1,ref:eit_2,ref:is_1,ref:is_2}.

Before showing a detailed discussion on non-equilibrium radiation in optically thin matters, let us summarize the point of radiation theory in optically thick matters: 
The collisionless nature of photons denotes that, when photons are in vacuum space in which no matter except photons exists, any dissipative flux never arises in the gas of photons (e.g. see \S63 in Landau-Lifshitz's textbook~\cite{ref:ll.statistical}). 
Hence, the traditional theory of radiative energy transfer~\cite{ref:stellar} has been applied to a mixture of a radiation field with a matter such as a dense gas or other continuous medium. 
In the traditional theory, it is assumed that the medium matter is dense (optically thick) enough to ignore the vacuum region among constituent particles of the matter. 
Then, the successive absorptions and emissions of photons by constituent particles of medium matter make it possible to assume that the photons are as if in local equilibrium states whose temperatures equal those of local equilibrium states of the dense medium matter. 
Some extensions of this traditional (local equilibrium) theory to local non-equilibrium radiative transfer in \emph{optically thick} matters have already been considered by, for example, Udey-Israel~\cite{ref:rad.dense_2} and Fort-Llebot~\cite{ref:rad.dense_1} in the framework of EIT. 
In their formulations, the local non-equilibrium state of radiation at a spacetime point is determined with referring to the local non-equilibrium state of dense medium matter at the same point, and the successive absorptions and emissions of photons by constituent particles of medium matter mimics the dissipation for radiation field. 
Due to this mimic dissipation, the EIT's formalism becomes applicable to non-equilibrium radiation field in continuous medium matter~\cite{ref:rad.dense_1,ref:rad.dense_2}.

Then, consider a non-equilibrium radiation in optically thin matters: 
When the mean-free-path of photons is long and we can not neglect the effect of free streaming of photons, the notion of mimic dissipation becomes inappropriate, because photons in the free streaming do not interact with other matters. 
Then, the evolution of non-equilibrium radiation field with long mean-free-path can never be described in the framework of EIT, since the EIT is the theory designed for \emph{dissipative} fluids. 
This appears as the inconsistency of bilinear form of entropy production rate~\eref{eq:eit.bilinear} with non-equilibrium radiation in optically thin matter, which can be concretely explained with using the steady state thermodynamics for a radiation field~\cite{ref:noneq.rad_4}. 
The remaining of this section is for the explanation of such inconsistency.

%%%%%%%%%%%%%%%%%%%%
\subsection{Inconsistency of EIT with Optically Thin Radiative Transfer}

\begin{figure}[t]
 \begin{center}
 \includegraphics[height=25mm]{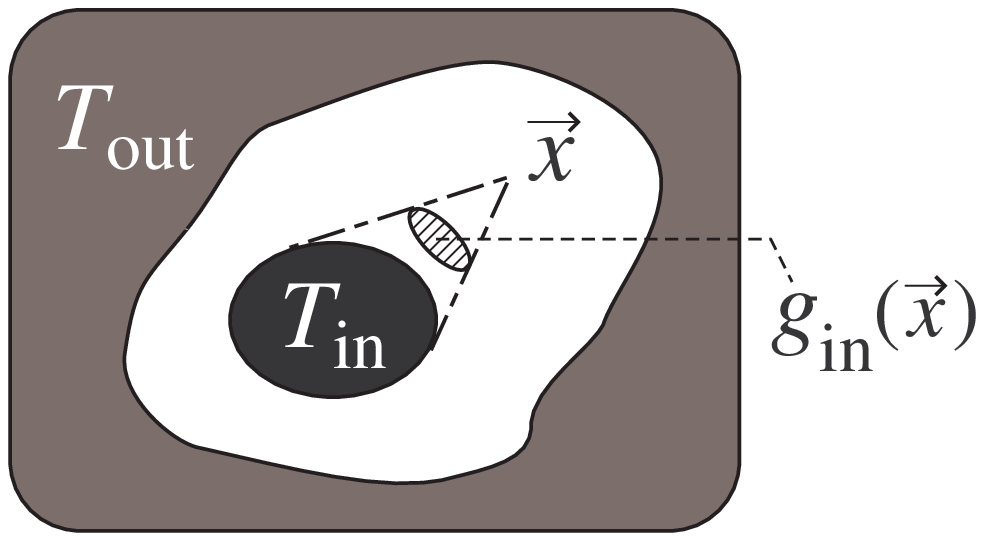}
 \end{center}
\caption{
{\small
A steady state of radiation field, which possesses a stationary (steady) energy flow in fixing temperatures~$T\ls{in}$ and~$T\ls{out}$. 
The non-equilibrium nature of this radiation field arises from the temperature difference. 
This is a typical model for non-equilibrium radiations in vacuum space or with long mean-free-path of photons. 
Even if the temperature difference is so small that the energy flux is weak and satisfies Eq.\eref{eq:eit.validity}, the time evolution of \emph{quasi-steady} processes of this radiation field can not be described by the EIT.
}
}
\label{fig:1}
\end{figure}

A significant case of radiation field in optically thin matters is the radiation field in \emph{vacuum} space, where the ``vacuum'' means that there exists no matter except a radiation field. 
As an example of a non-equilibrium radiation in vacuum space or with long mean-free-path of photons, let us investigate the system shown in Fig.\ref{fig:1}. 
For simplicity, we consider the case that any effect of gravity is neglected, and our discussion is focused on non-equilibrium physics without gravity. 
Furthermore, we approximate the speed of light to be infinity, which means that the size of the system shown in Fig.\ref{fig:1} is small enough.

In the system shown in Fig.\ref{fig:1}, a black body is put in a cavity. 
The inner and outer black bodies are individually in thermal equilibrium states, but those equilibrium states are different, whose equilibrium temperatures are respectively $T\ls{in}$ and~$T\ls{out}$. 
In the region enclosed by the two black bodies, there exists no matter except the radiation fields emitted by those black bodies. 
The photons emitted by the inner black body to a spatial point $\vec{x}$, which propagate through the shaded circle shown in Fig.\ref{fig:1}, have the temperature $T\ls{in}$. 
The other photons emitted by the outer black body have the temperature $T\ls{out}$. 
Therefore, although the inner and outer black bodies emit thermal radiation individually, the radiation spectrum observed at a point $\vec{x}$ is not thermal, since the spectrum has different temperatures according to the direction of observation. 
Furthermore, the directions and solid-angle around a point $\vec{x}$ covered by the photons emitted by inner black body, which is denoted by the shaded circle shown in Fig.\ref{fig:1}, changes from point to point in the region enclosed by the two black bodies. 
Hence, the radiation field is in \emph{local} non-equilibrium sates, whose radiation spectrum at one point is not necessarily the same with that at the other point. 
However, differently from Udey-Israel and Fort-Llebot theories~\cite{ref:rad.dense_2,ref:rad.dense_1}, there exists no reference non-equilibrium state of medium matter for the local non-equilibrium states of radiation shown in Fig.\ref{fig:1}, since the non-equilibrium radiation is in the vacuum region between two black bodies. 
The non-equilibrium radiation shown in Fig\ref{fig:1} is essentially different from those in optically thick medium. 
The system shown in Fig.\ref{fig:1}, which is composed of two black bodies and non-equilibrium radiation field between them, can be regarded as a representative toy model of radiative transfer with long mean-free-path of photons, and, when we focus on the non-equilibrium radiation field, it is a typical model of non-equilibrium radiation in vacuum space. 
Note that, when the temperatures $T\ls{in}$ and $T\ls{out}$ are fixed to be constant, the local non-equilibrium state of radiation at $\vec{x}$ has a stationary (steady) energy flux, $\vec{j}(\vec{x})$, due to the temperature difference. 
A non-equilibrium thermodynamic formulation has already been constructed for those \emph{steady} states of radiation field by one of present authors~\cite{ref:noneq.rad_4}.

When the steady non-equilibrium radiation system shown in Fig.\ref{fig:1} is compared with the heat conduction in continuum matters, one may expect that the energy flux in non-equilibrium radiation field, $\vec{j}$, corresponds to the heat flux in non-equilibrium continuum. 
Hence, according to the assumptions~2,~3 and~4, one may think it natural to assume that $\vec{j}$ is the dissipative flux which is the state variable characterizing non-equilibrium nature of steady states of radiation field in vacuum, and its entropy production rate is expressed by the bilinear form~\eref{eq:eit.bilinear}. 
However, from steady state thermodynamics for a radiation field~\cite{ref:noneq.rad_4}, it is concluded that the bilinear form of entropy production rate fails to describe an evolution of the system shown in Fig.\ref{fig:1}.

In order to review this fact, we need three preparations. 
First one is that the energy flux, $\vec{j}$, is not a state variable of the system shown in Fig.\ref{fig:1}. 
To explain it, recall that, in any thermodynamic theory, there exists a \emph{thermodynamic conjugate} state variable to any state variable. 
Therefore, if $\vec{j}$ is a state variable, there should exist a conjugate variable to $\vec{j}$. 
Here, for example, the temperature $T$ which is conjugate to entropy $S$ has a conjugate relation, $T = - \pdline{F}{S}$, where $F$ is the free energy. 
In general, thermodynamic conjugate variable can be obtained as a partial derivative of an appropriate thermodynamic functions such as internal energy, free energy, enthalpy and so on, which are related to each other by Legendre transformations. 
However, the following relation is already derived in the steady state thermodynamics for a radiation field~\cite{ref:noneq.rad_4},
\eqb
 \pd{F\ls{rad}}{j} = 0 \,,
\eqe
where $j = |\vec{j}|$ and $F\ls{rad}$ is the free energy of steady non-equilibrium radiation field. 
This denotes that thermodynamic conjugate variable to $\vec{j}$ does not exit, and $\vec{j}$ can never be a state variable of the system shown in Fig.\ref{fig:1}. 
Hence, if we apply the EIT's formalism to the steady non-equilibrium radiation field, the energy flux, $\vec{j}$, can not appear as a dissipative flux in the assumptions~2,~3 and~4.

Second preparation is to show the steady state entropy and two non-equilibrium state variables which are suitable to characterize the steady non-equilibrium radiation field instead of energy flux. 
The steady state thermodynamics for a radiation field~\cite{ref:noneq.rad_4} defines the density of steady state entropy, $s\ls{rad}(\vec{x})$, as
\eqb
\label{app:rad.srad}
 s\ls{rad}(\vec{x}) \defeq
  g\ls{in}(\vec{x})\,s\ls{eq}(T\ls{in}) + g\ls{out}(\vec{x})\,s\ls{eq}(T\ls{out}) \,,
\eqe
where $g\ls{in}(\vec{x})$ is the solid-angle of the shaded circle shown in Fig.\ref{fig:1} divided by $4 \pi$, $g\ls{out}(\vec{x})$ is the same for the remaining part of solid-angle around $\vec{x}$ satisfying $g\ls{in} + g\ls{out} = 1$ by definition, and $s\ls{eq}(T)$ is the density of equilibrium entropy of thermal radiation with equilibrium temperature $T$,
\eqb
 s\ls{eq}(T) \defeq \dfrac{16\,\sigma\ls{sb}}{3}\,T^3 \,,
\eqe
where $\sigma\ls{sb} \defeq \pi^2/60\hbar^3$ is the Stefan-Boltzmann constant. 
And, the other two state variables characterizing the steady states are a temperature difference and a kind of entropy difference, defined as
\seqb
\label{app:rad.tau_psi}
\eqab
 \tau\ls{rad} &\defeq& T\ls{in} - T\ls{out} \\
 \psi\ls{rad}(\vec{x}) &\defeq&
   g\ls{in}(\vec{x})\,g\ls{out}(\vec{x})\,
   \left[\, s\ls{eq}(T\ls{in}) - s\ls{eq}(T\ls{out}) \,\right] \,,
\eqae
\seqe
where we assume $T\ls{in} > T\ls{out}$ without loss of generality. 
If we apply the EIT's formalism to the system shown in Fig.\ref{fig:1}, the state variables characterizing the steady non-equilibrium radiation, $\tau\ls{rad}$ and $\psi\ls{rad}$, should be understood as the dissipative fluxes in the assumptions~2,~3 and~4.

Third preparation is the notion of \emph{quasi-steady process}. 
When the temperatures of inner and outer black bodies, $T\ls{in}$ and $T\ls{out}$, are kept constant, the non-equilibrium state of radiation field shown in Fig.\ref{fig:1} is stationary. 
However, if the whole system composed of two black bodies and radiation field between them is isolated from the outside of outer black body, then the whole system should relax to an equilibrium state in which the two black bodies and radiation field have the same equilibrium temperature. 
If the relaxation process proceeds so slowly, it is possible to approximate the time evolution of the slow relaxation as follows: 
The inner black body is in thermal equilibrium state of equilibrium temperature $T\ls{in}(t)$ at \emph{each} moment of time $t$ during the relaxation process. 
This means that the thermodynamic state of inner black body evolves on a sequence of equilibrium states in the space of thermodynamic states. 
This is the so-called quasi-static process in the ordinary equilibrium thermodynamics. 
Therefore, we can approximate the evolution of inner body by a quasi-static process. 
Also the evolution of outer black body is a quasi-static process on a sequence of equilibrium states which is different from that of inner black body's evolution. 
Then, at each moment of the slow relaxation process, the thermodynamic state of radiation field between two black bodies is regarded as a steady non-equilibrium state which possesses the state variables given in Eqs.\eref{app:rad.srad} and~\eref{app:rad.tau_psi}. 
This implies that, during the slow relaxation process of the whole system, the non-equilibrium radiation field evolves on a sequence of steady non-equilibrium states in the space of thermodynamic states. 
This is the quasi-steady process of non-equilibrium radiation field.

Given the above three preparations, we can show the inconsistency of EIT's formalism with non-equilibrium radiation fields in vacuum or with long mean-free-path of photons: 
To show it, we try to apply the EIT's formalism to the system shown in Fig.\ref{fig:1}, and will result in a failure. 
In order to satisfy the supplemental condition~1, which requires a sufficiently weak energy flux such as the inequality~\eref{eq:eit.validity} for dissipative matters, we consider the case with a sufficiently small temperature difference between two black bodies,
\eqb
 T\ls{in} = T\ls{out} + \delta T \,,
\eqe
where $\tau\ls{rad} = \delta T \ll e\ls{rad}$, and $e\ls{rad}$ is the energy density of steady non-equilibrium radiation field whose explicit definition~\cite{ref:noneq.rad_4} is not necessary here. 
Then, let us isolate the whole system composed of two black bodies and non-equilibrium radiation field from the outside of outer black body. 
The isolated whole system relaxes to an equilibrium state. 
Here, we focus our attention to the case of slow evolution of the relaxation process, which is regarded as a quasi-steady process. 
In this case, the time evolutions of temperatures are described by quantities, $T\ls{out}(t)$ and $\delta T(t)$, where $t$ is the time during the relaxation process. 
Under the assumption that the EIT's formalism works well for non-equilibrium radiation fields in vacuum or with long mean-free-path of photons, the time evolutions of $T\ls{out}(t)$ and $\delta T(t)$ should be determined by the EIT's formalism, in which the evolution equations of dissipative fluxes are obtained from the entropy production rate as reviewed in Sec.~\ref{sec:eqs.derivation}. 
Due to the requirement (4-b) in assumption~4, the entropy production rate of the relaxation process at time $t$ at point $\vec{x}$, $\sigma\ls{rad}(t,\vec{x})$, should be expressed by the bilinear form with using $\tau\ls{rad}$ and $\psi\ls{rad}$,
\eqb
\label{app:rad.sigma.bilinear}
 \sigma\ls{rad} = \tau\ls{rad}\,X_{\tau} + \psi\ls{rad}\,X_{\psi} \,,
\eqe
where $X_{\tau}$ and $X_{\psi}$ are respectively thermodynamic forces of $\tau\ls{rad}$ and $\psi\ls{rad}$. 
Then, because Eq.\eref{app:rad.sigma.bilinear} is a non-equilibrium equation of state for $\sigma\ls{rad}$, the supplemental condition~1 together with Eq.\eref{eq:eit.ass3} gives the relations,
\eqb
\label{app:rad.force_1}
 X_{\tau}(t,\vec{x}) = \lambda_{\tau}(t,\vec{x})\,\tau\ls{rad}(t) \quad,\quad
 X_{\psi}(t,\vec{x}) = \lambda_{\psi}(t,\vec{x})\,\psi\ls{rad}(t,\vec{x}) \,,
\eqe
where $\lambda_{\tau}$ and $\lambda_{\psi}$ are functions of fiducial equilibrium state variables which depend on $t$ and $\vec{x}$, and their non-negativity, $\lambda_{\tau} \ge 0$ and $\lambda_{\psi} \ge 0$, are obtained by the requirement (4-a) in assumption~4.

On the other hand, using the explicit form of $s\ls{rad}$ in Eq.\eref{app:rad.srad}, $\sigma\ls{rad}$ is given by
\eqab
\label{app:rad.sigma}
 \sigma\ls{rad}(t,\vec{x}) \defeq \pd{s\ls{rad}(t,\vec{x})}{t}
 =
  16\,\sigma\ls{sb}
    \left[\, g\ls{in}(\vec{x})\,T\ls{in}(t)^2\,\od{T\ls{in}(t)}{t}
           + g\ls{out}(\vec{x})\,T\ls{out}(t)^2\,\od{T\ls{out}(t)}{t} \,\right] \,,
\eqae
where $T\ls{in}(t) = T\ls{out}(t) + \delta T(t)$. 
Here, one may think that $\sigma\ls{rad}$ should include an entropy production due to the evolution of two black bodies. 
But, at the point $\vec{x}$ in the region enclosed by two black bodies, there exists only the non-equilibrium radiation field and the entropy production at that point should be due only to the non-equilibrium radiation field. 
Therefore, $\sigma\ls{rad}$ does not include contributions of entropies of black bodies.

In order to obtain the explicit forms of $X_{\tau}$ and $X_{\psi}$ from $\sigma\ls{rad}$ in Eq.\eref{app:rad.sigma}, we need to replace the pair of quantities $(\,T\ls{out}\,,\,\delta T\,)$ in Eq.\eref{app:rad.sigma} with the pair $(\,\tau\ls{rad}\,,\,\psi\ls{rad}\,)$. 
This replacement is carried out with the relations, $\tau\ls{rad} = \delta T$ and $\psi\ls{rad} \simeq 16\,\sigma\ls{sb}\,g\ls{in}\,g\ls{out}\,[\,T\ls{out}^2\,\delta T + T\ls{out}\,\delta T^2\,]$ up to $O(\delta T^2)$ due to supplemental condition~1. 
Then, Eq.\eref{app:rad.sigma} is rearranged to
\seqb
\eqb
\label{app:rad.sigma.rearrange}
 \sigma\ls{rad} =
 \dfrac{1}{D}\,\left[\,
 \tau\ls{rad}\,Z_{\tau} + \psi\ls{rad}\,Z_{\psi}
        + \tau\ls{rad}\,\psi\ls{rad}\,Z_{\tau\psi}
 \,\right] \,,
\eqe
where
\eqab
 Z_{\tau} &\defeq&
  16\,\sigma\ls{sb}\,g\ls{in}\,g\ls{out}\,\tau\ls{rad}^3\,
  \left(\,A - 2\,\sigma\ls{sb}\,g\ls{in}\,g\ls{out}\,\tau\ls{rad}^2 \,\right)\,
  \left[\, \pd{\psi\ls{rad}}{t}
         + 16\,\sigma\ls{sb}\,g\ls{in}\,g\ls{out}\,\tau\ls{rad}^2\,
           \od{\tau\ls{rad}}{t} \,\right] \\
 Z_{\psi} &\defeq& - A\,\psi\ls{rad}\,\od{\tau\ls{rad}}{t} \\
 Z_{\tau\psi} &\defeq&
  \left[\, A + 2\,\sigma\ls{sb}\,g\ls{in}\,g\ls{out}\,(1-4\,g\ls{out})\,
                  \tau\ls{rad}^2 \,\right]\,
  \left[\, \pd{\psi\ls{rad}}{t}
         + 16\,\sigma\ls{sb}\,g\ls{in}\,g\ls{out}\,\tau\ls{rad}^2\,
           \od{\tau\ls{rad}}{t} \,\right] \nonumber \\
  &&
  + 2\,\sigma\ls{sb}\,g\ls{in}\,g\ls{out}\,\tau\ls{rad}\,\psi\ls{rad}\,
    \od{\tau\ls{rad}}{t} \\
 D &\defeq&
  8\,g\ls{in}\,g\ls{out}\,\tau\ls{rad}^2\,A\,
  (A - 2\,\sigma\ls{sb}\,g\ls{in}\,g\ls{out}\,\tau\ls{rad}^2) \\
 A &\defeq&
  \sqrt{\sigma\ls{sb}\,g\ls{in}\,g\ls{out}\,\tau\ls{rad}\,
        (\,4\,\sigma\ls{sb}\,g\ls{in}\,g\ls{out}\,\tau\ls{rad}^3 + \psi\ls{rad}\,)}
 \,.
\eqae
\seqe
Therefore, with introducing an supplemental factor $\gamma\ls{rad}$, we obtain explicit forms of thermodynamic forces as
\eqb
\label{app:rad.force_2}
 X_{\tau} =
  \dfrac{1}{D}\,
  \left[\, Z_{\tau} + \gamma\ls{rad}\,\psi\ls{rad}\,Z_{\tau\psi} \,\right] 
 \quad,\quad
 X_{\psi} =
  \dfrac{1}{D}\,
  \left[\, Z_{\psi} + (1-\gamma\ls{rad})\,\tau\ls{rad}\,Z_{\tau\psi} \,\right] \,,
\eqe
where $\gamma\ls{rad}$ should be generally a function of fiducial equilibrium state variables which depend on $t$ and $\vec{x}$.

Hence, from Eqs.\eref{app:rad.force_1} and~\eref{app:rad.force_2}, we obtain two equations,
\eqb
\label{app:rad.eqs}
 D\,\lambda_{\tau}\,\tau\ls{rad}
  = Z_{\tau} + \gamma\ls{rad}\,\psi\ls{rad}\,Z_{\tau\psi} \quad,\quad
 D\,\lambda_{\psi}\,\psi\ls{rad}
  = Z_{\psi} + (1-\gamma\ls{rad})\,\tau\ls{rad}\,Z_{\tau\psi} \,.
\eqe
Here, recall that we are now seeking the evolution equations of $T\ls{out}(t)$ and $\delta T(t) = \tau\ls{rad}(t)$, which should be ordinary differential equations about time $t$. 
However, it is improbable to adjust the factors, $\gamma\ls{rad}(t,\vec{x})$, $\lambda_{\tau}(t,\vec{x})$ and $\lambda_{\psi}(t,\vec{x})$, so as to exclude the $\vec{x}$-dependence from Eq.\eref{app:rad.eqs} and yield the ordinary differential equations about $t$. 
Thus, evolution equations of $T\ls{out}(t)$ and $\delta T(t)$ can not be obtained in the framework of EIT. 
We conclude that the EIT's formalism fails to describe non-equilibrium radiation field in vacuum or with long mean-free-path of photons.

As explained above, EIT is not applicable to optically thin radiative transfer. 
However, a thermodynamic formulation for a stationary non-equilibrium radiation field in optically thin matters, which is different from EIT, has been constructed~\cite{ref:noneq.rad_4}. 
On the other hand, some efforts for describing non-equilibrium radiative transfer in optically thin astrophysical systems are now under the challenge, in which the so-called equation of radiative transfer is solved numerically under suitable approximations and assumptions (e.g. see a spatially tree-dimensional simulation for pseudo-Newtonian model by Kato, Umemura and Ohsuga~\cite{ref:noneq.rad_5}).
However, thermodynamic and/or hydrodynamic formulation of \emph{non-stationary} non-equilibrium radiation field in optically thin matters remains as an open and challenging issue.

%%%%%%%%%%%%%%%%%%%%%%%%%%%%%%%%%%%%%%%%%%%%%%%%%%%%%%%%%%%%%%%%%%%%%%%%%%%%%%%%%%%%%%%%%%%%%%%%%%%%
\section{Concluding Remark}
\label{sec:conc}

We have provided an axiomatic understanding of EIT, which is summarized in the basic assumptions and additional supplemental conditions shown in Sec.\ref{sec:ass}. 
Also the limit of EIT, which is not explicitly recognized in standard references of EIT~\cite{ref:eit_1,ref:eit_2,ref:is_1}, has been summarized in Sec.\ref{sec:rad}. 
Then, we end this manuscript with the following remark on a tacit understanding common to EIT and traditional laws of dissipations (Navier-Stokes and Fourier laws).

In the EIT, while thermodynamic state variables are treated via the second order dissipative perturbation as shown in supplemental condition~1, the dynamical variable (fluid velocity) is not subjected to the dissipative perturbation and remains as a function of independent thermodynamic state variables, $u^{\mu}(\ene,\Vne,q^{\alpha},\Pi,\sv\,^{\alpha\beta})$. 
This may imply that we have a tacit understanding as follows: 
Weak dissipative fluxes under supplemental condition~1 can raise a dissipative flow whose fluid velocity is essentially different from any fluid velocity of a perfect fluid's flow and can not be regarded as a perturbation of a perfect fluid's flow. 
If such a dissipative flow exits, then it can be described by the evolution equations~\eref{eq:eit.conservation} and~\eref{eq:eit.dissipation}. 
However, if some dissipative flow with weak dissipative fluxes is a perturbative flow of a perfect fluid's flow, then we should subject $u^{\mu}(\ene,\Vne,q^{\alpha},\Pi,\sv\,^{\alpha\beta})$ to the dissipative perturbation, $u^{\mu} = u^{\mu}\ls{(p)}(\eeq,\Veq) + \delta u^{\mu}$, where $\delta u^{\mu}$ is the velocity perturbation due to weak dissipative fluxes and $u^{\mu}\ls{(p)}$ is the flow of a back-ground perfect fluid determined independently of dissipative fluxes. 
In this case, the EIT's basic equations~\eref{eq:eit.conservation} and~\eref{eq:eit.dissipation} should also be rearranged into a perturbative form. 
(However, for the case that the perturbation $\delta u^{\mu}$ grows fast, such dissipative perturbation of fluid velocity would not work well.)

Note that the traditional laws of dissipations are extracted from EIT's basic equations~\eref{eq:eit.conservation} and~\eref{eq:eit.dissipation} by the limiting operation, $\tau\ls{h, b, s} \to 0$ and $\beta\ls{hb, hs} \to 0$ (vanishing relaxation time and interaction among dissipative fluxes).
This means that the traditional laws are also restricted to weak dissipations (in non-relativistic case).
Thus, the remark given in previous paragraph is also true of the traditional laws of dissipations.

%%%%%%%%%%%%%%%%%%%%%%%%%%%%%%%%%%%%%%%%%%%%%%%%%%%%%%%%%%%%%%%%%%%%%%%%%%%%%%%%%%%%%%%%%%%%%%%%%%%%
%\ack
\section*{Acknowledgement}

This work is supported by the grant of Daiko Foundation [No.9130], and partly by the Grant-in-Aid for Scientific Research Fund of the Ministry of Education, Culture, Sports, Science and Technology, Japan [Young Scientists (B) 19740149].

%%%%%%%%%%%%%%%%%%%%%%%%%%%%%%%%%%%%%%%%%%%%%%%%%%%%%%%%%%%%%%%%%%%%%%%%%%%%%%%%%%%%%%%%%%%%%%%%%%%%
\appendix
%%%%%%%%%%%%%%%%%%%%%%%%%%%%%%%%%%%%%%%%%%%%%%%%%%%%%%%%%%%%%%%%%%%%%%%%%%%%%%%%%%%%%%%%%%%%%%%%%%%%

%%%%%%%%%%%%%%%%%%%%%%%%%%%%%%%%%%%%%%%%%%%%%%%%%%%%%%%%%%%%%%%%%%%%%%%%%%%%%%%%%%%%%%%%%%%%%%%%%%%%
\section{Non-relativistic phenomenology of heat flux and viscosities}
\label{app:mc}

This appendix summarizes non-relativistic phenomenology of heat flux and viscosities. 
In this appendix, tensors are expressed as three dimensional quantities on three dimensional Euclidean space. 
The traditional laws of dissipations for heat flux and viscosities are summarized by the following three relations:
\seqb
\label{eq:mc.classic}
\eqab
\label{eq:mc.fourier}
 \mbox{Fourier law} &:& \vec{q} = - \lambda\,\vec{\nabla} T \\
\label{eq:mc.stokes}
 \mbox{Stokes law}  &:& \Pi = - \zeta\,\vec{\nabla}\cdot\vec{v} \\
\label{eq:mc.newton}
 \mbox{Newton law}  &:& \sv_{ij} = - 2 \eta\, \overset{\circ}{v}_{ij} \,,
\eqae
\seqe
where $T$ is the local equilibrium temperature, $\vec{q}$ and $\lambda$ are respectively the heat flux and heat conductivity, $\Pi$ and $\zeta$ are respectively the bulk viscosity and bulk viscous rate, $\sv_{ij}$ and $\eta$ are respectively the shear viscosity and shear viscous rate, and $\vec{v}$ and $\overset{\circ}{v}_{ij}$ are respectively the fluid velocity and shear velocity tensor defined as
\eqb
\label{eq:mc.vcirc}
 \overset{\circ}{v}_{ij} \defeq
 \partial_{(i} v_{j)} - \dfrac{1}{3}(\vec{\nabla}\cdot\vec{v})\,g_{ij} \,,
\eqe
where $g_{ij}$ is the metric of Euclidean space. 
Navier-Stokes equation is obtained by substituting Stokes and Newton laws into the non-relativistic version of Euler equation~\eref{eq:eit.eom}, in which the second and third terms in left-hand side and the terms including $q^{\mu}$ in right-hand side are the relativistic effects and disappear in non-relativistic case.

It should be emphasized that time derivatives of dissipative fluxes $\vec{q}$, $\Pi$ and $\sv_{ij}$ are not included in Eq.\eref{eq:mc.classic}. 
This means that the traditional laws of dissipations are phenomenological relations under the assumption that relaxation times of dissipative fluxes are zero. 
For example, it is assumed in the Fourier law that a non-stationary heat flux, $\vec{q}(t,\vec{x})$, relaxes \emph{instantaneously} to a stationary one, $\vec{q}(\vec{x})$, where $\vec{x}$ is the spatial coordinates. 
Therefore, the retarded effects of dissipative fluxes are ignored in the traditional laws of dissipations. 
This results in an infinitely fast propagation of perturbation of dissipative fluxes~\cite{ref:eit_1,ref:eit_2}. 
Hence, Eq.\eref{eq:mc.classic} can not describe dynamical dissipative phenomena whose dynamical time scale is comparable to the relaxation time scale of dissipative fluxes. 
This is the limit of Eq.\eref{eq:mc.classic} in either non-relativistic and relativistic cases.
Especially in relativistic cases, Eq.\eref{eq:mc.classic} violates the causality of dissipative phenomena.

In order to consider the retarded effects of dissipative fluxes, the simplest modification of Eq.\eref{eq:mc.classic} is to introduce the time derivative (Lagrange derivative) of dissipative fluxes.
This simple modification yields the phenomenological relations, which is called \emph{Maxwell-Cattaneo} laws~\cite{ref:eit_1,ref:eit_2}:
\eqb
\label{eq:mc.mc}
 \tauh\,\od{\vec{q}}{t} + \vec{q} \,=\, - \lambda\,\vec{\nabla} T \quad,\quad
 \taub\,\od{\Pi}{t} + \Pi \,=\, - \zeta\,\vec{\nabla}\cdot\vec{v} \quad,\quad
 \taus\,\od{\sv_{ij}}{t} + \sv_{ij} \,=\, - 2 \eta \,\overset{\circ}{v}_{ij} \,,
\eqe
where $\odline{Q}{t} = \pdline{Q}{t} + (\vec{v}\cdot\vec{\nabla}) Q$ is the Lagrange derivative in three dimensions, $\tau$'s are the relaxation times of dissipative fluxes. 
The finite speed of propagation of dissipative perturbation can be obtained by these laws. 
However, note that the three phenomenological relations in Maxwell-Cattaneo laws~\eref{eq:mc.mc} are independent each other. 
Therefore, the heating of fluid due to viscous flow and the occurrence of viscous flow due to heat flux can not be described by Eq.\eref{eq:mc.mc}. 
This means that the interactions among dissipative fluxes are not introduced in the Maxwell-Cattaneo laws. 
On the other hand, the EIT includes not only the finite propagation speed of dissipative effects but also the interactions among dissipative fluxes, which are represented by the coefficients in lists~\eref{eq:eit.coeff}. 
Furthermore, it is theoretically important to emphasize that, while the Maxwell-Cattaneo laws in Eq.\eref{eq:mc.mc} lack a systematic way how to add time derivative (Lagrange derivative) of dissipative fluxes, \emph{the framework of EIT gives the systematic method of introducing the Lagrange derivative of dissipative fluxes into their evolution equations}. 
The EIT is extendable to the other dissipation mechanisms such as diffusion among several components of fluid particles and electro-magnetic dissipation in plasma fluid~\cite{ref:eit_1,ref:eit_2}. 
Accepting EIT seems to be more promising than accepting Maxwell-Cattaneo laws.

%%%%%%%%%%%%%%%%%%%%%%%%%%%%%%%%%%%%%%%%%%%%%%%%%%%%%%%%%%%%%%%%%%%%%%%%%%%%%%%%%%%%%%%%%%%%%%%%%%%%
\section{Covariant derivative}
\label{app:cd}

On flat space, e.g. two dimensional Euclidean space $\mathbf{R}^2$, the derivative of a vector field $V^a$ ($a = 1, 2$ expressing the coordinates on $\mathbf{R}^2$),
\eqb
\label{eq:cd.pd}
\pd{V^a}{x^b} \defeq
 \lim_{\delta x^b \to 0}\dfrac{V^a(x^b+\delta x^b) - V^a(x^b)}{\delta x^b} \,,
\eqe
is defined with using the notion of ``parallel transport'' along $x^b$-axis. 
In Eq.\eref{eq:cd.pd}, the vector at point of $x^b+\delta x^b$ is parallel transported to the point of $x^b$ along $x^b$-axis, then the difference between the transported vector and the original vector at $x^b$ is calculated.

On curved spacetime, the ``parallel transport'' is defined so as to match with the ``curved shape'' of spacetime. 
The metric, which is the tensor field $g_{\mu\nu}$ of second rank, expresses the curved shape of the spacetime, where $\mu$ and $\nu$ denote the components of metric like $V^a$ of a vector in the previous paragraph. 
The length, $ds$, of the spacetime between infinitesimally near points $x^{\mu}$ and $x^{\mu}+dx^{\mu}$ is given as,
\eqb
\label{eq:cd.metric}
 ds^2 = g_{\mu\nu}\,dx^{\mu}\,dx^{\nu} \,,
\eqe
where left-hand side is the square of length $ds^2$, and $x^{\mu}$ denotes the coordinates on the spacetime. 
(For example, for two dimensional Minkowski spacetime, $ds^2 = -dt^2 + dx^2$ with ``rectangular coordinates'' $(t,x)$, which denotes the components of metric, $g_{\mu\nu} = {\rm diag.}(-1,1)$ where diag. means the ``diagonal matrix form''.) 
Then, in following the standard consideration in differential manifold, the covariant derivative of a vector field $W^{\mu}$ is given as,
\eqb
\label{eq:cd.cd.def}
 W^{\mu}_{\,\,\,;\nu} \defeq \lim_{\delta x^{\nu}\to 0}
 \dfrac{{\mathcal T}_{x^{\nu}}[W^{\mu}(x^{\nu}+\delta x^{\nu})] - W^{\mu}(x^{\nu})}
       {\delta x^{\nu}} \,,
\eqe
where ${\mathcal T}_{x^{\nu}}[W^{\mu}]$ is the parallel transport of $W^{\mu}$ to point $x^{\nu}$ along $x^{\nu}$-axis in curved spacetime. 
The parallel transport is explicitly expressed with using the metric, and results in
\eqb
\label{eq:cd.cd}
 W^{\mu}_{\,\,\,;\nu} = W^{\mu}_{\,\,\,,\nu} + \Gamma^{\mu}_{\nu\alpha}W^{\alpha} \,,
\eqe
where the Einstein's rule of contraction (the same indices appearing in upper and lower positions are summed, e.g. $W^{\mu}X_{\mu} = W^0X_0 + W^1X_1 + W^2X_2 + W^3X_3$ for coordinates $(x^0,x^1,x^2,x^3)$\,) is used, $\Gamma^{\mu}_{\nu\alpha}$ is the so-called Christoffel symbol given as
\eqb
 \Gamma^{\mu}_{\nu\alpha} \defeq
  \dfrac{1}{2}\,g^{\mu\beta}\,
  \Bigl(\,g_{\nu\beta\,,\alpha} + g_{\beta\alpha\,,\nu} - g_{\nu\alpha\,,\beta} \,\Bigr) \,,
\eqe
and the comma denotes the formal calculation of partial derivative, 
\eqb
 W^{\mu}_{\,\,\,,\nu} \defeq
 \lim_{\delta x^{\nu} \to 0}
 \dfrac{W^{\mu}(x^{\nu}+\delta x^{\nu}) - W^{\mu}(x^{\nu})}{\delta x^{\nu}} \,.
\eqe
In flat spacetime, $\Gamma^{\alpha}_{\mu\nu} = 0$, and the covariant derivative~\eref{eq:cd.cd} reduces to the partial derivative~\eref{eq:cd.pd}. 
Some important results of calculation in differential geometry are
\seqb
\eqab
 W_{\mu\,;\,\nu} &\defeq& g_{\mu\alpha}W^{\alpha}_{\,\,\,;\nu}
  = W_{\mu\,,\,\nu} - \Gamma^{\beta}_{\mu\nu}W_{\beta}\quad
 (\, W_{\mu} \defeq g_{\mu\alpha}W^{\alpha} \,) \\
 Y^{\mu\nu}_{\quad;\lambda} &=&
 Y^{\mu\nu}_{\quad,\lambda}
  + \Gamma^{\mu}_{\lambda\alpha}Y^{\alpha\nu} + \Gamma^{\nu}_{\lambda\alpha}Y^{\mu\alpha}\\
 Y^{\mu}_{\,\,\,\nu\,;\,\lambda} &\defeq& g_{\nu\alpha}Y^{\mu\alpha}_{\quad;\lambda} =
 Y^{\mu}_{\,\,\,\nu\,,\lambda}
  + \Gamma^{\mu}_{\beta\lambda}Y^{\beta}_{\,\,\,\nu}
  - \Gamma^{\beta}_{\nu\lambda}Y^{\mu}_{\,\,\,\beta}\\
 Y_{\mu\nu\,;\,\lambda} &\defeq& g_{\mu\alpha}Y^{\alpha}_{\,\,\,\nu;\lambda} =
 Y_{\mu\nu\,,\lambda}
  - \Gamma^{\beta}_{\mu\lambda}Y_{\beta\nu}
  - \Gamma^{\beta}_{\nu\lambda}Y_{\mu\beta} \,,
\eqae
and the metric is invariant under covariant derivative,
\eqb
 g_{\mu\nu\,;\lambda} = 0 \,.
\eqe
\seqe

%%%%%%%%%%%%%%%%%%%%%%%%%%%%%%%%%%%%%%%%%%%%%%%%%%%%%%%%%%%%%%%%%%%%%%%%%%%%%%%%%%%%%%%%%%%%%%%%%%%%
%\section*{References}

\end{document}